\documentclass[journal]{IEEEtran}

\usepackage{graphicx,color}
\usepackage{dcolumn}
\usepackage{amssymb,amsmath,amsfonts,bm, bbm}
\usepackage{amsfonts,bm, bbm}
\usepackage{graphicx}
\usepackage{textcomp}
\usepackage{amsthm}
\usepackage{mathtools}
\usepackage[ruled,vlined,linesnumbered]{algorithm2e}
\usepackage{enumitem}
\usepackage{physics}
\usepackage[export]{adjustbox}
\usepackage[normalem]{ulem}
\mathtoolsset{showonlyrefs=true}
\usepackage{cite}

\theoremstyle{plain}

\usepackage[dvipsnames]{xcolor}
\usepackage{booktabs}

\newcommand{\mc}{\mathcal}

\usepackage{hyperref}

\usepackage{mathrsfs}

\usepackage{soul}

\title{Iterative Decoding of Stabilizer Codes under Radiation-Induced Correlated Noise}

\begin{document}

\author{%
Anuj K. Nayak,
Paul G. Baity,
Peter J. Love,
Nicholas Jeon,
Byung-Jun Yoon,\\
Adolfy Hoisie, and
Lav R. Varshney\thanks{AKN and LRV are with the Department of Electrical and Computer Engineering and the Coordinated Science Laboratory, University of Illinois Urbana-Champaign, Urbana, IL, USA  (e-mail: anujk4@illinois.edu, varshney@illinois.edu).}
\thanks{PGB, PJL, BJY, AH, and LRV are with Brookhaven National Laboratory, Upton, NY, USA.}
\thanks{PJL is with the Department of Physics and Astronomy, Tufts University, Medford, MA, USA.}
\thanks{NJ and BJY are with the Department of Electrical and Computer Engineering, Texas A\&M University, College Station, TX, USA.}
\thanks{LRV is with the AI Innovation Institute, Stony Brook University, Stony Brook, NY, USA.}
\thanks{This work was supported in part by the Laboratory for Physical Sciences through Strategic Partnership Project No. EAOC0206928.}
\thanks{
This research used the Delta advanced computing and data resource which is supported by the National Science Foundation (award OAC 2005572) and the State of Illinois. Delta is a joint effort of the University of Illinois Urbana-Champaign and its National Center for Supercomputing Applications.}
}

\maketitle

\begin{abstract}
Fault-tolerant quantum computation demands extremely low logical error rates, yet superconducting qubit arrays are subject to radiation-induced correlated noise arising from cosmic-ray muon-generated quasiparticles. The quasiparticle density is unknown and time-varying, resulting in a mismatch between the true noise statistics and the priors assumed by standard decoders, and consequently, degraded logical performance. We formalize joint noise sensing and decoding using syndrome measurements by modeling the QP density as a latent variable, which governs correlation in physical errors and syndrome measurements. Starting from a variational expectation--maximization approach, we derive an iterative algorithm that alternates between QP density estimation and syndrome-based decoding under the updated noise model. Simulations of surface-code and bivariate bicycle quantum memory under radiation-induced correlated noise demonstrate a measurable reduction in logical error probability relative to baseline decoding with a uniform prior. Beyond improved decoding performance, the inferred QP density provides diagnostic information relevant to device characterization, shielding, and chip design. These results indicate that integrating physical noise estimation into decoding can mitigate correlated noise effects and relax effective error-rate requirements for fault-tolerant quantum computation.
\end{abstract}

\section{Introduction}
In superconducting quantum chips, high-energy particles from cosmic radiation, such as muons and gamma rays, can deposit energy in the substrate and generate QP bursts that induce correlated errors across multiple qubits, posing a significant challenge for fault-tolerant quantum computation (FTQC). Although the effects of ionizing radiations can be partially mitigated passively through methods such as enhanced shielding, backside metallization, and underground operation to reduce event frequency \cite{vepsalainen2020impact}, even a single cosmic-ray hit can cause chip-wide failure during long computations. The performance of quantum error correction (QEC) codes under these conditions is therefore critically dependent on the ability to model and mitigate correlated noise. 

In current QEC settings, such radiation-induced events are treated as blackouts or are implicitly absorbed into the logical error rate without explicit mitigation. QP-induced correlated errors using chip-wide “prepare-and-measure” protocols have been characterized \cite{mcewen2022resolving}. Within code design, hierarchical QEC codes that combine surface-code and quantum LDPC layers \cite{pattison2025hierarchical}, as well as interleaving techniques for toric codes \cite{trinca2022new}, have been proposed to handle burst-like correlated errors. These studies indicate the need for active, chip-level detection and suppression mechanisms to complement passive mitigation approaches. One can actively suppress correlated errors directly during decoding by tracking QP densities. Developing decoding algorithms tailored for radiation-induced correlations is important to improve QEC performance.

In classical coding theory, correlated noise processes have long been analyzed as channels with memory. The Gilbert-Elliott model formalized burst-error behavior in binary channels \cite{elliott1963estimates}, and subsequent studies on storage and communication systems investigated correlated and bursty error patterns \cite{johnson2009burst}. In magnetic storage, wireless communication, and high-speed interconnects, correlated noise due to intersymbol interference (ISI) and other physical effects has been mitigated through whitening filters, cyclic prefix with frequency domain reference signals, or interleaving \cite{andrews1997theory}. Further, iterative decoding methods such as for turbo codes \cite{hinton2002turbo} and LDPC codes \cite{hou2001performance} established probabilistic message-passing frameworks that handle correlated errors effectively. These suggest adapting classical approaches to quantum systems, in which cosmic-ray-induced quasi-particle bursts may be modeled as temporally and spatially correlated error events---a quantum counterpart to a channel with memory.

\begin{figure}
    \centering
    \includegraphics[width=0.95\linewidth]{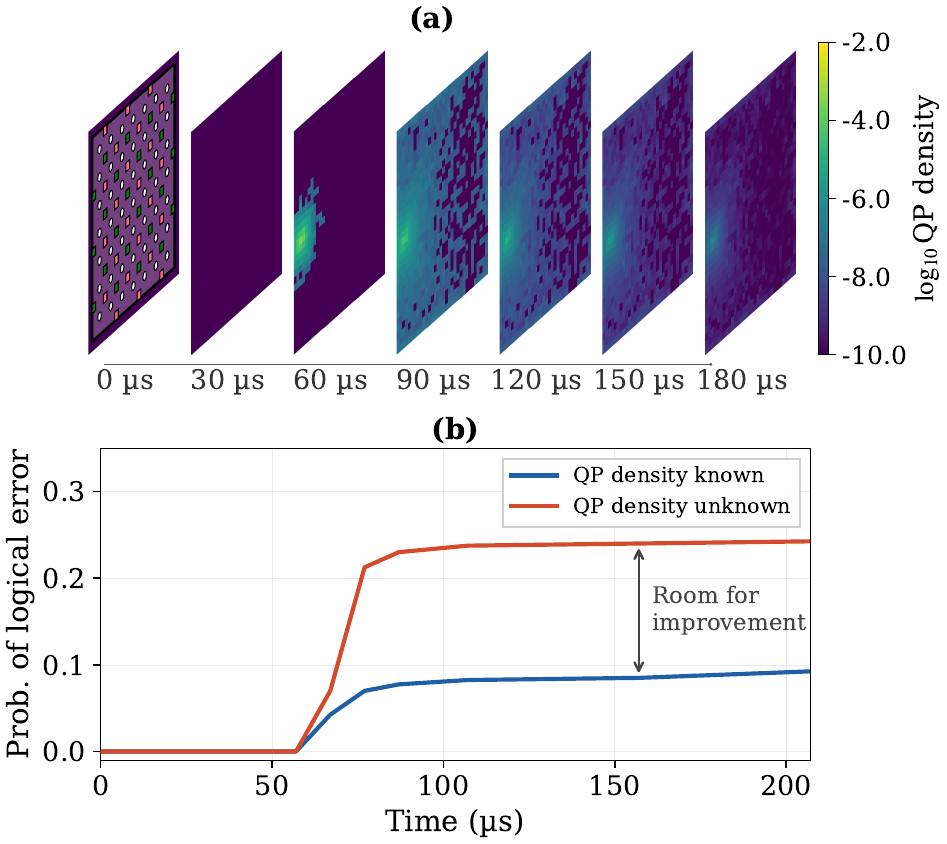}\vspace{-2mm}
    \caption{(a) Distance-7 surface code layout on a $40 \rm{mm} \times 40 \rm{mm}$ superconducting chip (shown at time slice $0~\mu s$) along with time varying QP density evolution heatmap; (b) Logical error probability (PLE) of decoder with perfect knowledge of instantaneous QP density and of the same decoder with uniform prior. The gap in PLE suggests room for improvement.}
    \label{fig:motivation_figure}
\end{figure}

Fig.~\ref{fig:motivation_figure} illustrates the impact of a muon strike on a $40\,\text{mm} \times 40\,\text{mm}$ superconducting chip hosting a distance-7 rotated surface code with a regular qubit layout. A \emph{genie} decoder with perfect knowledge of instantaneous QP density achieves a logical error probability (PLE) of about $9\%$, whereas an identical decoder operating under a fixed uniform prior only reaches roughly $24\%$, an approximately $2.7\times$ PLE degradation at $150$ \textmu s. The syndrome trajectory (sequence of stabilizer measurement outcomes collected at every QEC cycle) carries statistical signatures of spatially and temporally correlated errors, and may itself be sufficient to track QP density without auxiliary sensing hardware. We show this performance gap can largely be closed by jointly inferring the quasiparticle density and the most likely error configuration directly from the syndrome trajectory.

This paper develops techniques for joint estimation of QP dynamics and decoding via probabilistic inference on extended factor graphs. These methods improve decoding performance by modeling latent QP density within the inference process. The joint framework provides valuable estimates of QP density evolution alongside improved logical error rates in superconducting qubit systems under correlated noise.

\section{Background}
\subsection{Radiation-Induced Correlated Noise}

A high-energy cosmic ray or muon impact on a superconducting qubit chip causes an abrupt elevation in quasiparticle density, inducing transient degradation of qubit coherence \cite{vepsalainen2020impact}. Due to phonon-mediated coupling between distant qubits through the substrate, a single muon impact can induce spatially correlated multi-qubit error bursts that conventional surface or concatenated codes cannot mitigate efficiently \cite{martinis2021saving}.

The QP density time evolution is governed by the following differential equation \cite{yelton2024modeling}.
\begin{equation}
\tfrac{d x}{dt} = -r x^2 - s x + g(t),   
\label{eqn:qp_density_evolution_nonlin_ODE}
\end{equation}
where $x$ is the QP density at a specific spatial coordinate in a superconducting chip, $r$ is the QP recombination rate, $s$ is the QP trapping rate, and $g(t)$ is the QP generation term.

\subsection{Quantum Error Correction}

\subsubsection{Stabilizer Codes}

An $[[n,k]]$ stabilizer code encodes $k$ logical qubits into $n$ data qubits and is defined by an abelian subgroup
$\mathcal{S}=\langle S_1,\ldots,S_{n_a}\rangle$ of the Pauli group $\mathcal{P}_n$, with $-I\notin\mathcal{S}$ and $n_a=n-k$.
The codespace is the joint $+1$ eigenspace of all elements of $\mathcal{S}$ \cite{Gottesman1997}.
We consider a circuit-level realization with $n$ data qubits and $n_a$ ancilla qubits, for a total of
$N=n+n_a$ physical qubits.
Each stabilizer generator $S_i$ is measured using an associated ancilla qubit $a_i$ via a Clifford circuit
$U_i$ satisfying $U_i^\dagger Z_{a_i} U_i = S_i$.

Syndrome extraction is performed over $T$ stabilizer-measurement cycles indexed by $t=1,\ldots,T$.
Each cycle consists of $\mathcal{T}$ parallel rounds of entangling Clifford gates indexed by
$\tau=1,\ldots,\mathcal{T}$, interleaved with single-qubit gates, acting on the $N$ physical qubits,
followed by projective measurement of all ancilla qubits.
The measurement outcomes at the end of cycle $t$ define the syndrome vector
$S_t\in\mathbb{F}_2^{n_a}$.
Pauli faults propagate through the layered Clifford circuits by conjugation, and the resulting mapping
from fault locations to the syndrome trajectory $\{S_t\}_{t=1}^T$ is linear over $\mathbb{F}_2$ and fully
determined by the circuit geometry, gate ordering, and parallelization structure.

As an example, for the rotated distance-$7$ surface code, a single stabilizer-measurement cycle consists of $\mathcal{T}=7$ sequential rounds: two rounds of Hadamard gates applied to ancilla qubits to interchange $X$- and $Z$-type measurements, four rounds of nearest-neighbor entangling Clifford gates (CNOTs) implementing the stabilizer couplings in a fixed schedule, and a final ancilla measurement round. The code encodes $k=1$ logical qubit into $n=2d^2-1=97$ data qubits with $n_a=48$ ancilla qubits arranged on a $7\times7$ rotated lattice. Each cycle produces a syndrome vector $S_t\in\mathbb{F}_2^{48}$, and Pauli faults at any space--time location $(i,\tau,t)$ propagate through the seven-round circuit by Clifford conjugation, inducing detection events across neighboring stabilizers and adjacent cycles according to the surface-code geometry and gate ordering.

\section{Model}
\subsection{QP Density Evolution}

We linearize QP density evolution \eqref{eqn:qp_density_evolution_nonlin_ODE}. Let the latent QP densities be tracked over a grid of resolution $L \times W$ sites, and let $N$ be the total number of physical qubits occupying a subset of sites on the grid. We model radiation impact and QP poisoning as a linear dynamical system with latent QP field $X_t \in \mathbb{R}^{J}$ (with $J = N$). The latent field evolves as:
\begin{equation}
X_t = A_t X_{t-1} + \gamma C_t,
\end{equation}
where $A_t$ is the transition matrix and $C_t \in \mathbbm{Z}^{J}_+ \cup \{0\}$ is the spatiotemporally sparse QP injection vector. We do not model recombination since $r x^2$ is relatively small \cite{yelton2024modeling}. The time-dependent transition matrix is expressed as
$A_t = I - (sI + \kappa L) \, \Delta t$, where $\kappa$ is the scalar diffusion constant, $s$ is the QP trapping rate, $L$ is the graph Laplacian of dimension $J \times J$ that governs the spatial coupling and the diffusion dynamics of QPs across the chip. Take $L$ to be the random-walk normalized Laplacian of a Gaussian-weighted graph with adjacency
$W_{ij}=\exp\!\left(-\|\boldsymbol{\ell}_i-\boldsymbol{\ell}_j\|^2/(2\sigma^2)\right)\mathbf{1}\{j\in\mathcal{N}_\rho(i)\}$,
where $\boldsymbol{\ell}_i$ and $\boldsymbol{\ell}_j$ denote the spatial coordinates. Here, $\mathcal{N}_\rho(i)$ denotes the physical qubits within a radius of $\rho$ from $i$ and $\sigma$ is the kernel width. After symmetrization $W\leftarrow(W+W^\top)/2$, the Laplacian is $L=I-D^{-1}W$ with $D=\mathrm{diag}(W\mathbf{1})$. The operator $\kappa L$ thus corresponds to a discrete diffusion, with $\kappa$ controlling the effective diffusion coefficient.

\subsection{Circuit Noise Model}

\label{subsec:noise_model}

We assume the radiation-induced qubit error is the only source of physical noise in the stabilizer circuit and that the noise is Pauli-twirled. Noise is correlated when the underlying QP density profile is unknown. Conversely, the errors are conditionally independent (but not identically distributed) across physical qubits when instantaneous QP densities are known. Thus, given the instantaneous QP density, the noise model is circuit-level, time-varying Pauli noise with independent but non-identically distributed (i.n.i.d.) errors on all qubits. Within a given round, all physical qubits experience the same duration $\Delta t$ (equal to the gate, idle, or readout time, whichever is longest), but the resulting Pauli error probabilities differ across qubits according to the local QP density $x_{t,i}$. Given QP density $x$ at a specific location and time, the Pauli (twirled) error probabilities are
\begin{align}
\label{eqn:pxyz}
p_X(x) &= p_Y(x) = \tfrac{1}{4} \left(1 - e^{-\Delta t/T_1(x)}\right), \mbox{ and}\\
p_Z(x) &= \tfrac{1}{4}\left(1 + e^{-\Delta t/T_1(x)} - 2 e^{-\Delta t/T_2(x)} \right),
\end{align}
where $\Delta t$ is the gate, idle duration or readout duration, and
$T_1/T_2$ times are computed from QP density following \cite{Baity2026}:
\begin{align}
T_1(x) &=
\left[
2 \sqrt{2\,\frac{\Delta_{\mathrm{Al}}}{h}\, \frac{\omega_{01}}{2\pi}}\, x + 1 / T_1^{b}
\right]^{-1},\\
T_2(x) & =\left[\frac{1}{2\,T_1(x)}+
\frac{E_C}{h}\frac{x^2}{\pi}\,
\exp\!\left(\frac{1}{2}\, W_0\left(\frac{4\pi}{ x^2}\right)\right)
\right]^{-1},
\label{eqn:T1_T2_xqp}
\end{align}
where $W_0(\cdot)$ denotes the principal branch of the Lambert-W function, and the nominal values of the parameters are $\Delta_{\mathrm{Al}} = 191 \,\mu\mathrm{eV}$, $E_C/h = 400\,\mathrm{MHz}$, $\omega_{01}/(2\pi)=5\,\mathrm{GHz}$, and $T_1^{b}=100\,\mu\mathrm{s}$. 

We denote the function that maps QP density to Pauli error probabilities as $\Psi : \mathbb{R}_+ \to [0,1]^2: x \mapsto (p_X(x),\, p_Z(x))$, where $p_Y = p_X$ and $p_I = 1 - 2p_X - p_Z$. Since each component is strictly monotone on $x \geq 0$, the component-wise inverses 
$\Psi_X^\dagger : [0,1] \to \mathbb{R}_+:p_X \mapsto x$, and $\Psi_Z^\dagger : [0,1] \to \mathbb{R}_+:p_Z \mapsto x$ are well-defined:
\begin{align}
    \Psi_X^\dagger(p) &= \frac{-\frac{1}{\Delta t}\ln(1 - 4p) - \frac{1}{T_1^b}}{2\sqrt{2\,\frac{\Delta_{\mathrm{Al}}}{h}\frac{\omega_{01}}{2\pi}}}, \mbox{and}\\
    \Psi_Z^\dagger(p) &= \{x \geq 0 : p_Z(x) = p\},
    \label{eqn:psi_dag}
\end{align}
where $\Psi_X^\dagger$ is closed-form and $\Psi_Z^\dagger$ is computed numerically by standard scalar root-finding.

\begin{figure*}
    \centering
    \includegraphics[width=0.95\linewidth]{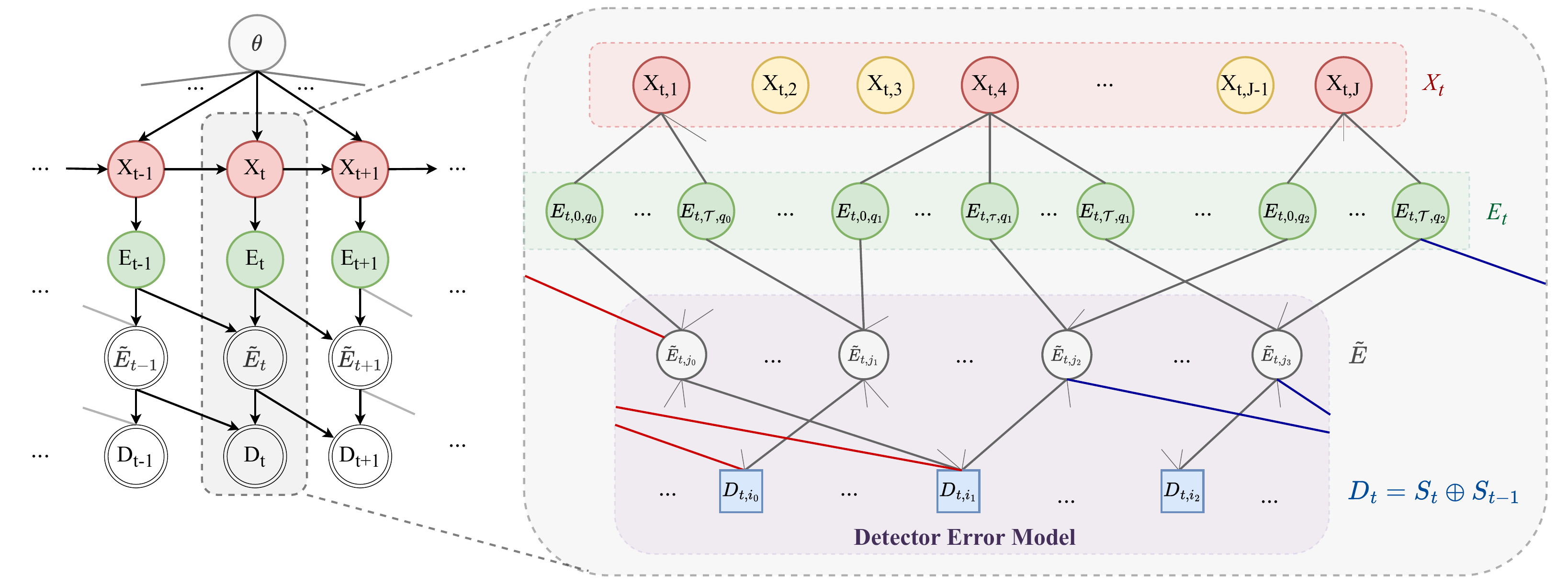}\vspace{-2mm}
    \caption{Graphical model coupling QP density evolution, circuit faults, error mechanisms and detection events.}
    \label{fig:graphical_model}
\end{figure*}

\subsection{Detector Error Model}
\label{subsec:dem}
A detection event associated with an ancilla qubit $i$ is the change in stabilizer outcomes between consecutive cycles, i.e., $D_{t,i} = S_{t,i} \oplus S_{t-1, i}\in\mathbb{F}_2$, with $S_{0,i}$ initialized to zero. A Pauli fault $E_{t,\tau, q} \in \{I, X, Y, Z\}$ is defined as the insertion of a Pauli operator at a specific circuit location, corresponding to a qubit index $q$, cycle $t$, and after round $\tau$. Each $E_{t,\tau, q}$ causes detection pattern $D_{1:T} = [D_{t,i}]_{t, i} = \phi(E_{t,\tau, q}) \in \mathbb{F}_2^{n_a T}$. For a given detection event pattern, $D^{(j)}_{1:T}$, the set of Pauli faults causing it is defined as an error mechanism, i.e., $\tilde{\mc{E}}_j \triangleq \{E_{t,\tau,q} \mid \phi(E_{t,\tau,q}) = D^{(j)}_{1:T}\} $  and is indexed by $j \in \{1, \ldots, n_{\tilde{E}}\}$, where $n_{\tilde{E}}$ is the number of unique error mechanisms. The detector error model (DEM) is a bipartite graph (Tanner graph) that describes the map between the set of error mechanisms $\{\tilde{\mc{E}}_j\}_j$ and the set of detectors $\{\mc{D}_{t,i}\}_{t,i}$, see Fig.~\ref{fig:graphical_model}. Equivalently, DEM is specified by a sparse binary matrix $H$ whose columns index error mechanisms and whose rows index detectors $\mc{D}_{(t,i)}$, such that $H_{(t,i),(t',j)}=1$ if an error mechanism $\tilde{\mc{E}}_{t', j}$ flips detector $\mc{D}_{(t,i)}$ \cite{Fowler2012, gidney2021stim}. The weights of the error mechanisms are approximated as the sum of the constituent Pauli fault probabilities from the noise model described in Sec.~\ref{subsec:noise_model}, $p(\tilde{\mc{E}}_{t,j}) = \sum_{E \in \tilde{\mc{E}}_{t,j}} p(E)$ (similar to the decoder in \cite{bravyi2024high}). Decoding then reduces to a graph-based inference problem on the DEM graph, yielding a most-likely equivalence class of Pauli faults consistent with the observed detector pattern.

Each error mechanism affects detectors in at most two consecutive cycles, say $D_{t}, D_{t+1}$. This allows the error mechanisms to be indexed by $(t,j)$ as $\tilde{\mc{E}}_{t,j}$, where $j \in \{1, \ldots, n_{\tilde{E}_t}\}$ and $n_{\tilde{E}_t}$ is the number of error mechanisms in cycle $t$. Moreover, a Pauli fault $E_{t,\tau,q}$ is contained in either $\tilde{\mc{E}}_{t,j}$ or $\tilde{\mc{E}}_{t+1,j}$ for some $j$. This spatio-temporal coupling induced by circuit-level noise is reminiscent of spatially coupled codes like convolutional codes. While spatial coupling is determined by the underlying Tanner graph topology of the stabilizer code, the temporal coupling has a constraint-length of 2 cycles (see Appendix~\ref{apndx:pcm}).

\subsection{Probabilistic Graphical Model}
\label{subsec:pgm}

Figure \ref{fig:graphical_model} shows generative model from QP density profile to detection sequence. Let $\theta$ denote the parameters that govern the QP density profile (e.g., the diffusion parameter $\kappa$). The generative model induces the Markov chain \(\theta \leftrightarrow X_{1:T} \leftrightarrow E_{1:T} \leftrightarrow \tilde{E}_{1:T} \leftrightarrow D_{1:T}\). We assume $\{X_t \mid \theta\}_{t=1}^T$ is a Markov process. The random variable  $\tilde{E}_{t,j} = \mathbbm{1}_{\{\tilde{\mc{E}}_{t,j}\}}$ indicates the occurrence of at least one of the constituent Pauli faults in the error mechanism $\tilde{\mc{E}}_{t, j}$, and $\tilde{E}_{t} = [\tilde{E}_{t,j}]_j \in \mathbbm{F}_2^{n_{\tilde{E}_t}}$ is the corresponding Bernoulli random vector. Similarly, $D_t$ is also a function of $(\tilde{E}_{t-1}, \tilde{E}_t)$.

Let us zoom into a specific cycle $t$ (see Figure~\ref{fig:graphical_model}). An edge exists between $X_{t,i}$ and a Pauli fault $E_{t,\tau,q}$ if qubit $q$ is situated at site $i$. Moreover, the Pauli fault $E_{t,\tau,q}$ is conditionally independent of all other faults given QP density $X_{t,i}$, i.e., $E_{t,\tau,q} \perp (E_{1:t-1}, E_{t+1:T}, E_{t,\sim \tau,\sim q}) | X_{t,i}$. Edges between Pauli faults and error mechanisms is defined by $\Phi(\cdot)$, as discussed in \ref{subsec:dem}. The spatio-temporal coupling between error mechanisms and detectors is governed by the stabilizer code and circuit-level noise model.

\subsection{Decoding}
\subsubsection{Belief Propagation}
Given the DEM Tanner graph $H$ and the error-mechanism prior probabilities $\{p(\tilde{\mathcal{E}}_{t,j})\}_{t,j}$ computed from the noise model $\Psi$ (Sec.~\ref{subsec:noise_model}), decoding reduces to marginal inference on a sparse binary factor graph. Belief propagation (BP) solves this by iteratively passing log-likelihood ratio (LLR) messages between variable nodes $\tilde{E}_{t,j}$ and check nodes $\mathcal{D}_{t,i}$ \cite{Gallager1962LDPC, Kschischang2001FactorGraphs, Forney2001FactorGraphs}. Let $\lambda_{j \to i}^{(\ell)}$ and $\mu_{i \to j}^{(\ell)}$ denote the variable-to-check and check-to-variable LLR messages at iteration $\ell$, respectively. The message update rules are:
\begin{align}
    \lambda_{j\to i}^{(\ell)} &= \lambda_j^{(0)} + \sum_{i' \in \mathcal{N}(j) \setminus i} \mu_{i' \to j}^{(\ell-1)}, \\
    \mu_{i \to j}^{(\ell)} &= 2 \tanh^{-1}\!\left(\prod_{j' \in \mathcal{N}(i) \setminus j} \tanh\!\tfrac{\lambda_{j' \to i}^{(\ell)}}{2}\right),
\end{align}
where $\lambda_j^{(0)} = \log\frac{1 - p(\tilde{\mathcal{E}}_{t,j})}{p(\tilde{\mathcal{E}}_{t,j})}$ is the channel LLR initialized from the prior, and $\mathcal{N}(\cdot)$ denotes the set of neighbors in $H$. After $L_{\mathrm{BP}}$ iterations, the posterior LLR for each error mechanism is $\hat{\lambda}_j = \lambda_j^{(0)} + \sum_{i \in \mathcal{N}(j)} \mu_{i \to j}^{(L_{\mathrm{BP}})}$, and a hard decision is made as $\hat{\tilde{E}}_{t,j} = \mathbbm{1}\{\hat{\lambda}_j < 0\}$. The estimated syndrome is checked as $\hat{D} = H\tilde{E}_{1:T} \pmod{2}$; if $\hat{D} = D_{1:T}$, BP has converged to a valid decoding. The time-varying, i.n.i.d.\ error probabilities induced by the spatially inhomogeneous QP density profile enter BP solely through the initialization of channel LLRs $\{\lambda_j^{(0)}\}$, making BP naturally compatible with the QP-aware noise model.

\subsubsection{Post-Processing: Ordered Statistics Decoding}
BP is known to underperform on codes with short cycles in their Tanner graphs, a characteristic shared by topological codes such as the surface code \cite{MacKayNeal1997NearShannon, PoulinChung2008SparseQuantumBP}. When BP fails to converge, i.e., $\hat{D} \neq D_{1:T}$, ordered statistics decoding (OSD) is applied as a post-processor \cite{FossorierLin1995OSD, Roffe2020DecodingLandscape, PanteleevKalachev2023DegenerateCodes}. OSD uses the soft information produced by BP: the error mechanisms are sorted in decreasing order of reliability $|\hat{\lambda}_j|$, and a systematic form of $H$ is constructed by Gaussian elimination over $\mathbb{F}_2$ using the most reliable columns as the information set. Let $\mathcal{I} = \{j_1, \ldots, j_{n_a}\}$ index the pivot columns selected during elimination, forming a full-rank submatrix $H_\mathcal{I}$. OSD-$\omega$ then enumerates all binary test vectors $\mathbf{e}_\mathcal{I} \in \mathbb{F}_2^{n_a}$ of Hamming weight at most $\omega$ over the information set, computes the corresponding codeword $\hat{\tilde{E}} = H_\mathcal{I}^{-1}(D_{1:T} \oplus H_{\bar{\mathcal{I}}}\mathbf{e}_{\bar{\mathcal{I}}})$ for each candidate, and selects the candidate of highest posterior probability under the BP LLRs. The combined BP+OSD decoder thus pairs the computational efficiency of message passing with the reliability guarantees of soft-decision algebraic post-processing, yielding a decoder well-suited to the structured, inhomogeneous noise arising from spatially correlated QP poisoning.

\section{Problem Formulation}
Our objective is to jointly infer the latent spatiotemporal QP field $X_{1:T} \in \mathbb{R}^{J \times T}$ and its governing parameters $\theta$ from the observed detection events $D_{1:T} \in \mathbb{F}^{n_a T}_2$. Specifically, we maximize the log-joint, $\log p(X_{1:T}, \theta, D_{1:T})$. Introducing a variational distribution $q(E_{1:T}, \tilde{E}_{1:T})$ and applying a standard variational decomposition (see Appendix~\ref{apndx:variational_decomposition}) yields
\begin{align}
\label{eqn:log_map_objective}
\log& ~ p(X_{1:T}, \theta, D_{1:T}) \\
&= \log p(X_{1:T}, \theta)
+ \mathbb{E}_{q(\tilde{E}_{1:T})}
\!\left[
\log p(\tilde{E}_{1:T} \mid X_{1:T})
\right] \nonumber \\
&\quad + H(q(\tilde{E}_{1:T})) + \mathrm{KL}\!\left(
q(\tilde{E}_{1:T})
\,\|\, p(\tilde{E}_{1:T} \mid D_{1:T}, X_{1:T})\right) \nonumber \\
&\quad + \text{degeneracy residual.}
\label{eqn:elbo_basic}
\end{align}
The first three terms on the right side constitute the variational lower bound, and the KL term vanishes when $q$ matches the true posterior $p(\tilde{E}_{1:T} \mid D_{1:T}, X_{1:T})$. However, this posterior is intractable: BP decoders produce only error mechanism-wise marginals under the Bethe approximation \cite{yedidia2003understanding}, which deviate from the true posterior, and the error mechanism prior $p(\tilde{E}_{1:T} \mid X_{1:T})$ is itself approximated by a mean-field factorization $p_m(\tilde{E}_{1:T} \mid X_{1:T}) = \prod_{j} p(\tilde{E}_j \mid X_{1:T})$. It is therefore instructive to factor both the Bethe posterior $p_b(\tilde{E}_{1:T} \mid D_{1:T}, X_{1:T})$ and the mean-field prior $p_m$ explicitly into the decomposition. As shown in Appendix~\ref{apndx:variational_decomposition}, this yields
\begin{align}
\log& ~p(X_{1:T}, \theta, D_{1:T}) \\
&= \log p(X_{1:T}, \theta)
+ \mathbb{E}_{q(\tilde{E}_{1:T})}
\!\left[
\log p_m(\tilde{E}_{1:T} \mid X_{1:T})
\right] \nonumber \\
& + H(q(\tilde{E}_{1:T})) + \mathrm{KL}\!\left(
q(\tilde{E}_{1:T})
\,\|\, p_b(\tilde{E}_{1:T} \mid D_{1:T}, X_{1:T})\right) \nonumber \\
& + \text{degeneracy} + \text{mean-field approx.} + \text{algorithmic error},
\label{eqn:elbo_bp}
\end{align}
where the three residual terms respectively capture errors from the degeneracy of the error mechanism representation, the mean-field factorization of the prior, and the deviation of the Bethe posterior from the true posterior (see Appendix~\ref{apndx:variational_decomposition} for their explicit forms). Treating these residuals as irreducible terms, and maximizing the first three terms motivates a variational EM algorithm that alternates between inferring $q^{(k)}(\tilde{E}_{1:T})$ and updating $(X_{1:T}, \theta)$.

\subsubsection{Variational E-step (Decoding)}
At iteration $k$, the E-step sets the variational distribution to the Bethe posterior evaluated at the current QP estimate $X_{1:T}^{(k)}$:
\begin{equation}
q^{(k)}(\tilde{E}_{1:T}) := p_b\!\left(\tilde{E}_{1:T}\mid D_{1:T}, X_{1:T}^{(k)}\right).
\label{eqn:e_step_temporal}
\end{equation}
Operationally, this corresponds to running the BP decoder with QP-dependent channel priors $p(\tilde{E}_{1:T} \mid X_{1:T}^{(k)})$, which yields soft marginals over error mechanisms.

\subsubsection{M-step (Sensing)}

The M-step updates the QP field and parameters by maximizing the tractable terms of the ELBO 
with $q$ fixed at $q^{(k)}$:
\begin{align}
X_{1:T}^{(k+1)}, \theta^{(k+1)} := &\underset{X_{1:T},\, \theta}{\mathrm{argmax}} \;
\log p(X_{1:T}, \theta) \nonumber \\
&+ \mathbb{E}_{q^{(k)}(\tilde{E}_{1:T})}
\!\left[\log p_m(\tilde{E}_{1:T} \mid X_{1:T})\right].
\label{eqn:m_step_temporal}
\end{align}
This step fuses the spatiotemporally correlated information in the syndrome-weighted error 
mechanism posteriors with the QP dynamics prior $p(X_{1:T}, \theta)$, effectively propagating decoding information into the QP field estimate.

\section{Algorithms}
Building upon the Markov chain structure and independent error mechanism assumptions established in Section~\ref{subsec:pgm}, we now exploit the spatiotemporal conditional independence of the probabilistic graphical model to derive a computationally tractable joint inference procedure. The QP density evolution satisfies the Markov factorization
\begin{equation}
p(X_{1:T}, \theta) = p(\theta)\, p(X_0 \mid \theta) \prod\limits_{t = 1}^{T} 
p(X_t \mid X_{t-1}, \theta),
\end{equation}
and the mean-field approximation used in belief propagation yields a factorized 
variational distribution over error mechanisms, $q(\tilde{E}_{1:T}) = \prod_{j} q(\tilde{E}_j)$. Under these assumptions, the joint log-likelihood decomposes as 
\begin{equation}
\mc{L}(X_{1:T}, \theta) 
= \mc{L}_{\rm{qp}}(X_{1:T}, \theta) + \mc{L}_{\rm{err}}(X_{1:T}) + \mc{L}_{\rm{diff}}(\theta),
\end{equation}
where $\mc{L}_{\rm{qp}}$ captures the QP diffusion dynamics, $\mc{L}_{\rm{err}}$ couples the QP density to the observed syndromes through the error mechanism probabilities, and $\mc{L}_{\rm{diff}}$ encodes the prior on the diffusion parameter $\kappa$. These terms are given by
\begin{equation}
\mc{L}_{\rm{qp}}(X_{1:T}, \theta) = \sum\limits_{t=1}^{T} \log p(X_t \mid X_{t-1}, \theta)
\end{equation}

\begin{align}
    \mc{L}_{\rm{err}}(X_{1:T}) = \sum\limits_{t,j} & q(\tilde{E}_{t,j}) \log p_m(\tilde{E}_{t,j} \mid X_{1:T}) \\ &+ (1 - q(\tilde{E}_{t,j})) \log (1-p_m(\tilde{E}_{t,j} \mid X_{1:T})).
\end{align}
The term $\mc{L}_{\rm{err}}$ is a binary cross-entropy between the BP posterior $q(\tilde{E}_{t,j})$ and the prior probability of error mechanism $\tilde{\mc{E}}_{t,j}$ predicted by the current QP density estimate. The conditional probability $p_m(\tilde{E}_{t,j} \mid X_{1:T})$ is approximated by summing over the contributing single Pauli faults,
\begin{align}
    p_m(\tilde{E}_{t,j}\mid X_{1:T}) \approx \sum\limits_{\substack{E_{t,\tau,q}:\Phi(E_{t, \tau, q}) \in \tilde{\mc{E}}_{t,j}}} p(E_{t, \tau, q} | X_{t,q}),
\end{align}
where $p(E_{t,\tau,q} \mid X_{t,q}) \triangleq p_{E_{t,\tau,q}}(X_{t,q})$ as defined in~\eqref{eqn:pxyz}. This approximation~\cite{bravyi2024high} is accurate when the probability of two or more simultaneous Pauli faults within a single error mechanism is negligible relative to single-fault contributions.

We further adopt the following simplifying assumptions.

\subsection{Algorithm 1}

We further make the following simplifying assumptions.
\begin{enumerate}
    \item The QP injection vector $C_t$ is absorbed into the stochastic fluctuations 
    $W_t$, so that $\theta = \{\kappa\}$.
    \item The QP density evolves log-normally: $\log X_t \mid \log X_{t-1}, \kappa \sim 
    \mc{N}\!\left(\log(A(\kappa) X_{t-1}),\, Q\right)$, where $Q$ is the process noise 
    covariance.
    \item A Beta prior is placed on the diffusion parameter: $\kappa \sim \mathrm{Beta}(a, b)$.
\end{enumerate}

Under assumption~2, the QP dynamics term becomes a weighted least squares penalty in 
log-space,
\begin{align}
  \mc{L}_{\rm{qp}}(X_{1:T}, \kappa) = -\frac{1}{2} \sum_{t=1}^{T} ||  \log X_t - \log\left(A(\kappa) X_{t-1}\right) ||^2_{Q^{-1}},
\end{align}
and the prior on $\kappa$ contributes a log-Beta regularization term,
\begin{equation}
    \mc{L}_{\rm{diff}}(\kappa) = (a-1)\log\,(\kappa) + (b-1) \log\,(1-\kappa).
\end{equation}

The temporal coupling structure of the DEM provides additional computational structure. For repeated cycles of syndrome measurement under circuit-level noise, the syndrome at cycle $t$ depends only on error mechanisms spanning at most two adjacent cycles,
\begin{equation}
    D_{t} = [H^{(t-1)}_2 | H^{(t)}_0 | H^{(t)}_1] \, [\tilde{E}^{(0)}_{t-1} | \tilde{E}^{(0)}_{t} | \tilde{E}^{(1)}_{t}]^T
\end{equation}
where $H^{(t)}_0$, $H^{(t)}_1$, $H^{(t-1)}_2$ are the parity check submatrices corresponding to faults localized within round $t$, spanning rounds $t$ and $t+1$, and spanning rounds $t-1$ and $t$, respectively (see Appendix~\ref{apndx:pcm} for more details). This temporal locality means that $\nabla_{Z_{t,i}} \mc{L}_{\rm{err}}$ receives contributions only from error mechanisms $\tilde{\mc{E}}_{t',j}$ whose Tanner graph neighborhood includes qubit $i$ at time $t$, yielding a sparse gradient. This sparsity motivates a \emph{sliding window} approach~\cite{gong2024toward} to joint decoding and QP density estimation, in which inference is performed over a window of $T_w$ cycles at a time, with the window advancing by a stride of $t_s$ cycles. Let $t_l$ and $t_h = t_l + T_w - 1$ denote the start and end of the current window $w$, with $H(w)$, $\tilde{E}(w)$, and $D(w)$ denoting the corresponding submatrices and vectors. Within each window, we alternate between a belief propagation (BP) decode step, which updates the error mechanism posteriors $q(\tilde{E}_{t,j})$ given the current QP density estimate, and a gradient ascent step that refines $Z_{t_l:t_h}$ and $\kappa$ to maximize $\mc{L}_{\rm{qp}} + \mc{L}_{\rm{err}} + \mc{L}_{\rm{diff}}$. The offline (batch) algorithm is recovered as the special case $T_w = T$. The full procedure is summarized in Algorithm~\ref{algo:decode_sense_sliding_window}.

The algorithm is governed by three loop parameters: $W = \lceil (T - T_w) / t_s \rceil + 1$ is the total number of sliding windows; $K$ is the number of outer BP--gradient ascent iterations per window; and $M$ is the number of gradient ascent steps per BP update. At each outer iteration $k$, the map $\Psi$ converts the log-QP density $Z_{t_l:t_h}^{(k)} \in \mathbb{R}^{J \times T_w}$ into per-qubit Pauli error probabilities $P_{m,\,t_l:t_h}^{(k)} \in [0,1]^{3 \times J \times \mathcal{T} T_w}$, which are then aggregated into error mechanism probabilities by $\Phi$. BP uses these error mechanism priors and the detector outcomes, yielding marginals $q^{(k)} \in [0,1]^{n_{\tilde{E}(w)}}$, where $n_{\tilde{E}(w)}$ is the number of error mechanisms in window $w$. These marginals inform $M$ gradient ascent steps that refine $Z_{t_l:t_h}^{(k)}$ and $\theta^{(k)}$. After $K$ outer iterations, $\texttt{PostProc}(q^{(K)})$ applies OSD to the final BP marginals to produce the hard-decision error vector $\tilde{E}(w) \in \{0,1\}^{n_{\tilde{E}(w)}}$ (MWPM may be substituted for surface codes). $\texttt{UpdateError}(\tilde{E}(w))$ commits the decoded error mechanisms from the leading $t_s$ cycles of window $w$ to $\tilde{E}^*$, leaving the trailing $T_w - t_s$ cycles for refinement in the next window. To provide a warm-start for the next window, $\texttt{Propagate}(Z_{t_l:t_h}^{(K)},\,\theta^{(K)})$ extrapolates $Z_{1:T} \in \mathbb{R}^{J \times T}$ forward from $t_h$ to $t_h + t_s$ through the mean recursion
\begin{equation}
    Z_{t_h+\tau}^{(1)} = \log A\!\left(\kappa^{(K)}\right) + Z_{t_h+\tau-1}^{(1)},
    \quad \tau = 1, \ldots, t_s,
\end{equation}
initialized at $Z_{t_h}^{(1)} = Z_{t_h}^{(K)}$.

\begin{algorithm}
\caption{Iterative Decoding and QP Density Estimation}
\label{algo:decode_sense_sliding_window}
\DontPrintSemicolon
\KwIn{$\{D_t\}_{t=1}^T$}
\KwOut{$X^*_{1:T}$, $\tilde{E}^*$}
\textbf{Initialize:} $Z_0$, $Z_{t_l:t_h}^{(1)}$, $\theta^{(1)}$, $t_l$, $t_s$, $t_h$\;
\For{$w = 1$ \KwTo $W$}{
    \For{$k = 1$ \KwTo $K$}{
        $P_{m,\, t_l:t_h}^{(k)} \leftarrow \Psi\!\left(\exp\!\left(Z_{t_l:t_h}^{(k)}\right)\right)$\;
        $q^{(k)} \leftarrow \mathrm{BeliefProp}\!\left(D_{t_l:t_h},\, \Phi\!\left(P_{m,\,t_l:t_h}^{(k)}\right)\right)$\;
        \For{$m = 1$ \KwTo $M$}{
            {\scriptsize $Z_{t_l:t_h}^{(k+1)} \leftarrow Z_{t_l:t_h}^{(k)} + \alpha \Bigl(\nabla_{Z_{t_l:t_h}^{(k)}}\mathcal{L}_{\mathrm{qp}} + \nabla_{Z_{t_l:t_h}^{(k)}}\mathcal{L}_{\mathrm{err}} \Bigr)$}\;
            {\scriptsize $\theta^{(k+1)} \leftarrow \theta^{(k)} + \beta \Bigl(\nabla_{\theta^{(k)}}\mathcal{L}_{\mathrm{qp}} + \nabla_{\theta^{(k)}}\mathcal{L}_{\mathrm{diff}} \Bigr)$}\;
        }
    }
    $\tilde{E}(w) \leftarrow \texttt{PostProc}(q^{(K)})$\;
    $D_{t_l:t_h} \leftarrow D_{t_l:t_h} \oplus H(w)\,\tilde{E}(w)$\;
    $Z_{t_h+1:\,t_h+t_s}^{(1)} \leftarrow \texttt{Propagate}\!\left(Z_{t_l:t_h}^{(K)},\, \theta^{(K)}\right)$\;
    $[\tilde{E}^*_{t_l:\,t_l+t_h-1,\,j}]_j \leftarrow \texttt{UpdateError}(\tilde{E}(w))$\;
    $X^*_{t_l:\,t_l+t_s} \leftarrow \exp\!\left(Z_{t_l:\,t_l+t_s}^{(K)}\right)$\;
    $t_l \leftarrow t_l + t_s$,\quad $t_h \leftarrow t_h + t_s$\;
}
\end{algorithm}

\subsection{Algorithm 2}

As an alternative to the gradient-based algorithm of the previous section, we now 
derive a recursive estimator for the QP density by casting the inference problem as 
nonlinear state estimation. The key idea is to introduce \emph{pseudo-measurements} 
that summarize the information about $X_t$ contained in the BP posteriors, allowing 
the error term $\mc{L}_{\rm{err}}$ to be approximated by a Gaussian likelihood 
amenable to Kalman filtering.

\paragraph{Pseudo-measurements}
The pseudo-measurement at qubit $i$ and time $t$ is defined as the average QP density 
inferred by inverting the Pauli error rate map $\Psi$ over all fault types and gate 
layers,
\begin{equation}
    Y_{t,i} = \frac{1}{n_{t,i}} \sum\limits_{\tau,\, E_{t,\tau,q}} 
    \Psi^\dagger_{E_{t,\tau,q}}\!\left(p_b(E_{t,\tau,q} \mid X_{t,i},\, D_{1:T})\right),
\end{equation}
where $\Psi_{E}^\dagger$ denotes the inverse of the Pauli fault probability ($E \in \{X, Y, Z\}$) and $n_{t,i}$ is a normalization count defined below. Since the true fault posterior 
$p_b(E_{t,\tau,q} \mid X_{t,i}, D_{1:T})$ is not directly available from the BP 
decoder, we approximate it by re-weighting the BP marginal $q^{(k)}(\tilde{E}_{t',j})$ 
by the ratio of the single-fault prior to the total error mechanism prior,
\begin{equation}
\hat{p}_b(E_{t,\tau,q} \mid X_{t,i}, D_{1:T}) = 
\frac{p(E_{t,\tau,q} \mid X^{(k)}_{t,i})}{p_m(\tilde{E}_{t',j} \mid X^{(k)}_{1:T})}\, 
q^{(k)}(\tilde{E}_{t',j}),
\end{equation}
where $\Phi(E_{t,\tau,q}) \in \tilde{\mc{E}}_{t',j}$. This approximation distributes the posterior probability of each error mechanism 
proportionally among its contributing Pauli faults according to their prior weights. 
At iteration $k$ of the EM procedure, the pseudo-measurement heuristic is
\begin{equation}
    Y^{(k+1)}_{t,i} = \frac{1}{n_{t,i}} 
    \sum\limits_{t',\tau,j} \hspace{-2mm}
    \sum\limits_{\substack{E_{t,\tau,q}:\\ \Phi(E_{t,\tau,q}) \in \tilde{\mc{E}}_{t',j}}}\hspace{-6mm}
    \Psi^\dagger_{E_{t,\tau,q}}\!\left(\hat{p}_b(E_{t,\tau,q} \mid X_{t,i},\, D_{1:T})\right),
\end{equation}
where $n_{t,i} = \lvert\{(\tau, j) : \Phi(E_{t,\tau,q}) \in \tilde{\mc{E}}_{t',j},\, 
\forall\, t'\}\rvert$ counts the total number of fault-mechanism pairs that involve 
qubit $i$ at time $t$.

\paragraph{Gaussian Measurement Model}
We model the pseudo-measurement noise as additive Gaussian in log-space,
\begin{equation}
    \log Y_t \mid \log X_t \;\sim\; \mc{N}(\log X_t,\; R),
\end{equation}
so that the error likelihood term becomes a Gaussian least-squares penalty,
\begin{equation}
    \mc{L}_{\rm{err}}(X_{1:T}) = -\frac{1}{2} \sum\limits_{t=1}^{T} 
    \|\log Y_t - \log X_t\|^2_{R^{-1}}.
\end{equation}
Combined with the log-normal QP dynamics of assumption~2, this yields the 
nonlinear state-space model
\begin{align}
   \log X_t &= \log \left( A(\kappa)\, X_{t-1}\right) + U_t, \\
   \log Y_t &= \log X_t + V_t,
\end{align}
where $U_t \sim \mc{N}(0, Q)$ and $V_t \sim \mc{N}(0, R)$, with isotropic covariances 
$Q = \sigma_Q^2\, \mathbbm{I}_K$ and $R = \sigma_R^2\, \mathbbm{I}_K$. The nonlinearity 
in the process equation arises from the log of the diffusion map $A(\kappa)$, which we 
linearize via a first-order Jacobian to obtain an extended Kalman filter (EKF). The 
predict and update equations of the EKF are given in 
Appendix~\ref{apndx:kalman_equations}.

\begin{algorithm}
\caption{Extended Kalman Filter}
\label{algo:kalman}
\DontPrintSemicolon
$P_{m, t_l:t_h} \leftarrow \Psi\!\left(\exp\left(Z_{t_l:t_h}\right)\right)$\;
$q \leftarrow \mathrm{BeliefProp}\!\left(D_{t_l:t_h}, \Phi(P_{m, t_l:t_h})\right)$\;
$Y_{t_l:t_h} \leftarrow \mathrm{PseudoMeas}\!\left(q, P_{m, t_l:t_h}\right)$\;
\For{$t = t_l$ \KwTo $t_h$}{
{\small
    $Z_{t|t-1},\, P_{t|t-1} \leftarrow \mathrm{EKFPredict}\!\left(Z_{t-1|t-1},\, P_{t-1|t-1},\, \theta\right)$\;
    $Z_{t|t},\, P_{t|t} \leftarrow \mathrm{EKFUpdate}\!\left(Z_{t|t-1},\, P_{t|t-1},\, Y_t\right)$\;
}
}
\end{algorithm}

The decoding and QP density estimation iterations in lines 3 through 10 in Algorithm~\ref{algo:decode_sense_sliding_window} are replaced by EKF updates in Algorithm~\ref{algo:kalman}. Here we treat $\theta$ as a fixed parameter.

\section{Results and Discussion}
\subsection{Simulation Setup}

We evaluate the proposed algorithms on QP density trajectories generated from G4CMP simulations~\cite{yelton2024modeling, Kelsey2023, Agostinelli2003, Allison2006, Allison2016} of a $40\,\mathrm{mm} \times 40\,\mathrm{mm}$ superconducting chip with thickness $525\,\mu\mathrm{m}$ and backside metallization. QP densities are discretized over a $32 \times 32$ spatial grid at $1\,\mu\mathrm{s}$ temporal resolution, and are treated as quasi-static within each stabilizer cycle.

\begin{figure}
    \centering
    \includegraphics[width=0.95\linewidth]{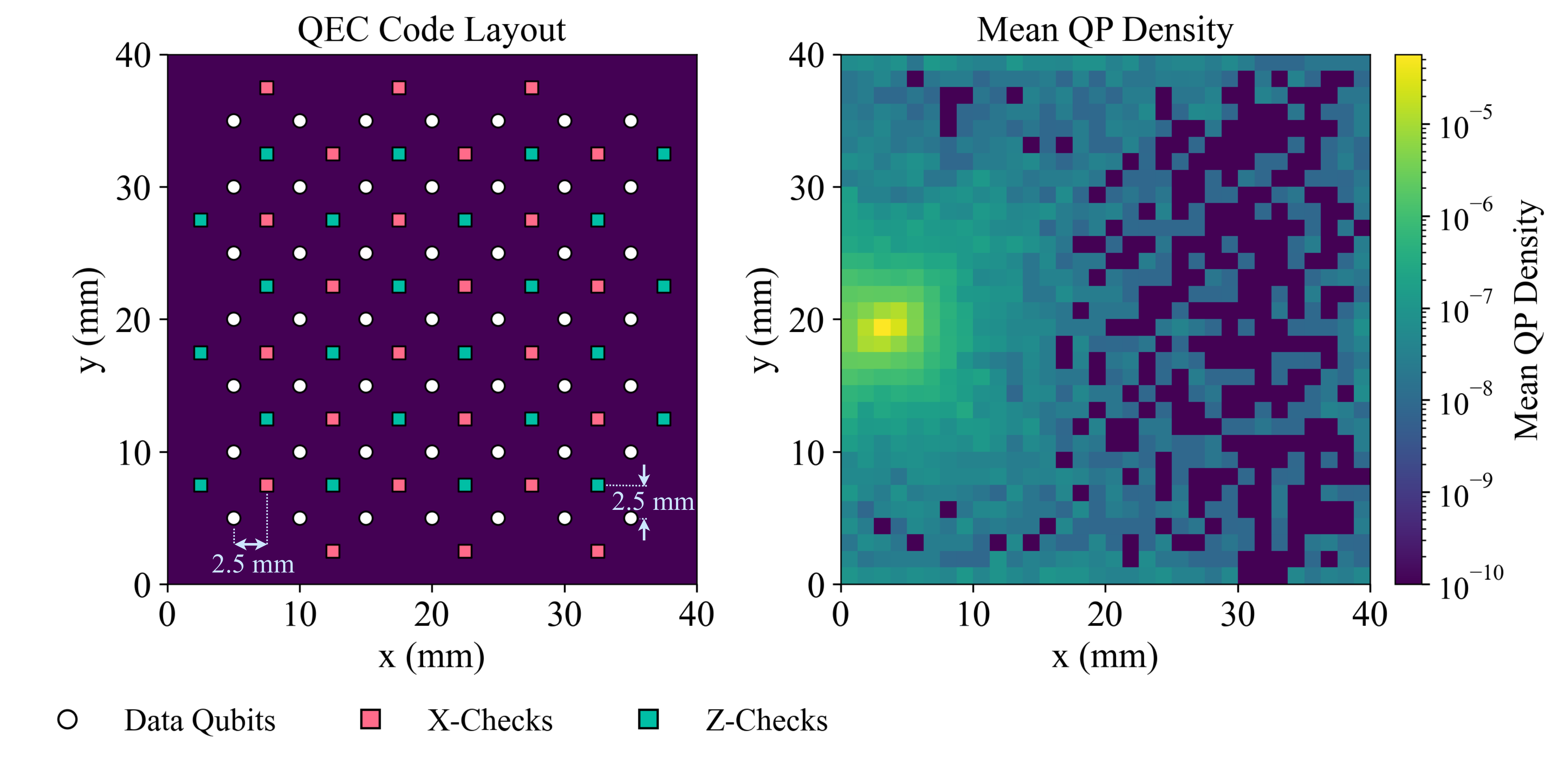}\vspace{-2mm}
    \caption{Distance 7 surface code layout (left) and time-averaged QP density of muon strike sample 58 (right).}
    \label{fig:code_layout_xqp_surface_xqp58}
\end{figure}

\begin{figure}
    \centering
    \includegraphics[width=0.95\linewidth]{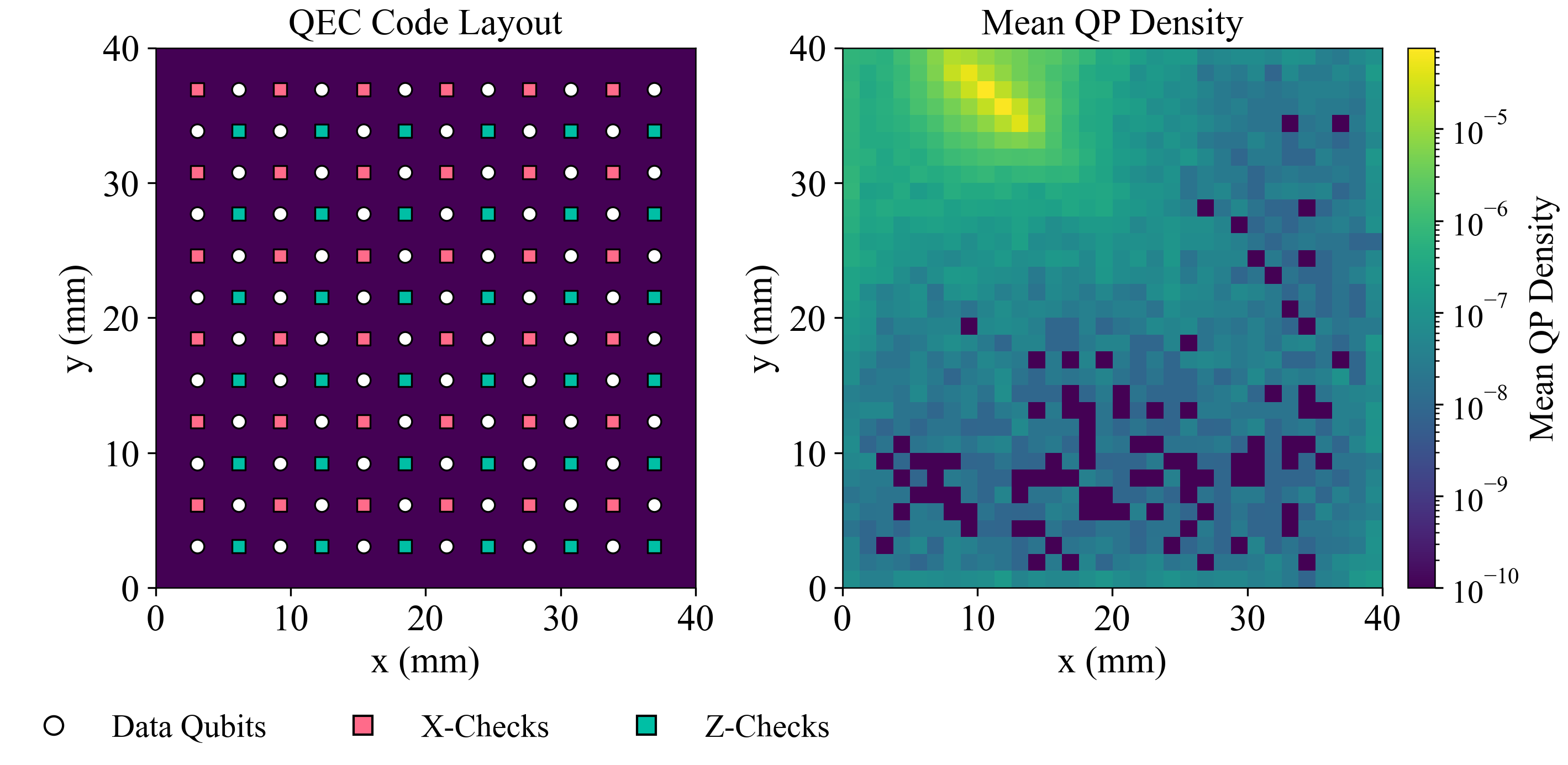}\vspace{-2mm}
    \caption{[[72, 12, 6]] BB qLDPC layout (left) and time-aveaged QP density of muon strike sample 53 (right).}
    \label{fig:code_layout_xqp_bb_qldpc_xqp53}
\end{figure}

Two stabilizer codes are evaluated: (i) a distance-7 rotated surface code with $n = 97$ data qubits, $n_a = 48$ ancilla qubits, and inter-qubit spacing of $2.5\,\mathrm{mm}$ in a staggered layout; and (ii) the $[[72,12,6]]$ bivariate bicycle (BB) qLDPC code with vertical and horizontal qubit spacings of $3.077\,\mathrm{mm}$. Physical qubit locations are arranged on a regular lattice, equidistant along both axes. The code layouts and representative time-averaged QP density fields for individual muon strike samples are shown in Figs.~\ref{fig:code_layout_xqp_surface_xqp58} and~\ref{fig:code_layout_xqp_bb_qldpc_xqp53}. Stabilizer circuit simulations are performed in \texttt{Stim}~\cite{gidney2021stim}. For both codes, gate durations follow the schedule of~\cite{google2023suppressing}: $\Delta t = 25\,\mathrm{ns}$ for Hadamard rounds (1 and 6), $\Delta t = 34\,\mathrm{ns}$ for CNOT rounds (2--5), and $\Delta t = 735\,\mathrm{ns}$ for syndrome measurement and reset (round 7), yielding a total cycle duration of $921\,\mathrm{ns}$.

The initial log-QP field $Z_0$ is set uniformly to $\log(10^{-8})$ at all sites. Algorithm~1 uses the Adam optimizer with a cosine learning rate schedule: linear warm-up to $0.5$, followed by cosine decay to $0.01$, with initial rate $0.01$. Model hyperparameters\footnote{In our simulations, we fix the parameter $\kappa$ obtained via Bayesian optimization, since we did not observe measurable gain with Beta prior and gradient ascent over $\kappa$.} $(\kappa, s, \sigma_Q^2)$ are selected per algorithm and code via Bayesian optimization using Optuna's TPE~\cite{akiba2019optuna}, minimizing the mean log-space maximum MSE between the estimated and G4CMP ground-truth QP density over 10 held-out muon samples (see Appendix~\ref{apndx:bayes_param_select} and Table~\ref{tab:bayes_best_params}).

Parameters of Algorithm~\ref{algo:decode_sense_sliding_window} are $T_w = 20$, $t_s = 10$ for the sliding-window variant and $T_w = T$ for the offline (batch) variant. Algorithm~2 uses $T_w = 2$, $t_s = 1$. Decoding uses 20 BP iterations (sum-product) followed by OSD-10. Five decoder configurations are compared: 1) \textit{Genie (offline)} with $T_w = T$ and priors set from ground-truth G4CMP QP densities. This serves as a PLE lower bound. 2) \textit{Algorithm~1 (offline)} with $T_w = T$, i.e., QP density estimated via gradient ascent over the full horizon. 3) \textit{Algorithm~1} with $(T_w=20,\,t_s=10)$, a sliding-window gradient-ascent decoder. 4) \textit{Algorithm~2} with $(T_w=2,\,t_s=1)$, an online EKF decoder. 5) \textit{Uniform} $(T_w=2,\,t_s=1)$: Decoder with a fixed uniform prior with QP density $10^{-8}$ at all sites and times, which serves as a PLE upper bound. All the configurations use 20 belief propagation iterations (sum-product algorithm) with OSD of 10 as post processing.

Burst errors adversely affect decoding within about 50 \textmu s of muon impact. Thereafter, the decoder is capable of tracking the subsequent errors, resulting in plateauing of PLE after 50 \textmu s. PLE reported at $t = 100$ \textmu s is sufficient to demonstrate the effectiveness of the iterative decoding algorithms upon a muon strike.

\subsection{Demonstrative Results}

\subsubsection{QP Density Estimation Accuracy.}
Fig.~\ref{fig:mse_boxplot} reports the log-space MSE between estimated and ground-truth QP density fields, aggregated across muon samples, for both codes. Algorithm~1 (offline) achieves the lowest median MSE in both cases (${\approx}0.70$ for the surface code, ${\approx}1.07$ for the BB-qLDPC code) with tight interquartile spread, consistent with its access to the full syndrome trajectory and unrestricted optimization horizon. The sliding-window variant of Algorithm~1 $(T_w=20,\,t_s=10)$ shows moderately higher median MSE (${\approx}0.93$ and ${\approx}1.17$, respectively) and a wider spread, reflecting the reduced temporal context available per window. Algorithm~2 $(T_w=2,\,t_s=1)$ exhibits substantially higher MSE (median ${\approx}2.31$ for the surface code, ${\approx}4.70$ for BB-qLDPC) with the largest spread across both codes, attributable to the highly compressed two-cycle window that provides limited syndrome context for QP inference per update step.

\subsubsection{Decoding Performance}

\begin{figure}
    \centering
    \includegraphics[width=0.9\linewidth]{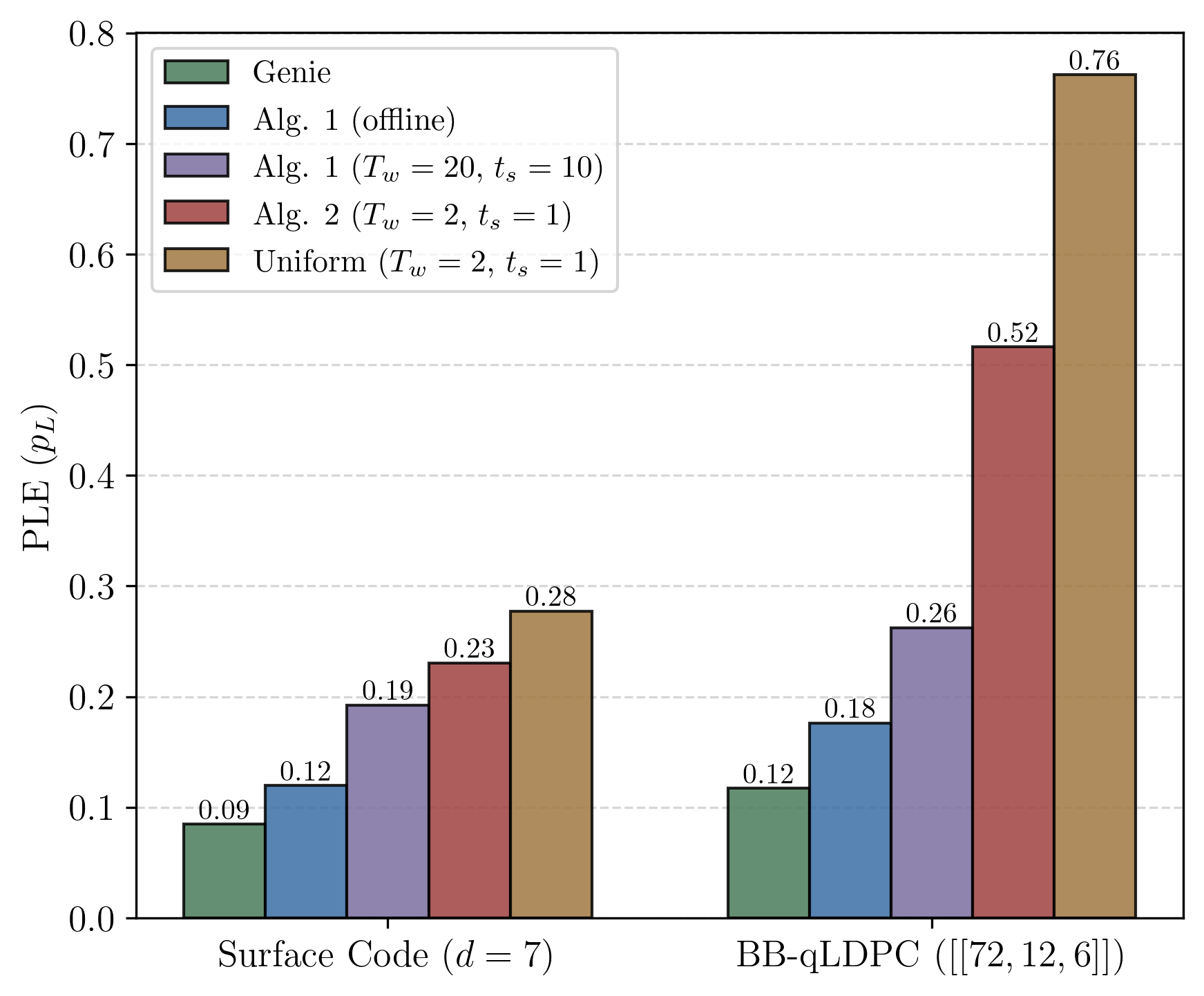}\vspace{-2mm}
    \caption{PLE at $100\,\mu s$ of distance 7 surface code and $[[72,12,6]]$ BB-qLDPC codes. Algorithm~\ref{algo:decode_sense_sliding_window} and \ref{algo:kalman} are compared with uniform prior baseline sliding window decoder (PLE upper bound), and full-horizon decoding with genie prior (PLE lower bound).}
    \label{fig:ple_bar_graph}
\end{figure}

Fig.~\ref{fig:ple_bar_graph} reports PLE at $t = 100$ \textmu s for all five configurations across both codes.

For the surface code ($d=7$), the uniform prior yields $p_L = 0.28$. The genie decoder achieves $p_L = 0.09$, establishing the performance achievable with perfect instantaneous QP knowledge. All three proposed configurations improve over the uniform baseline: Algorithm~1 (offline) reaches $p_L = 0.12$, the sliding-window variant achieves $p_L = 0.19$, and Algorithm~2 reaches $p_L = 0.23$. The offline variant narrows the gap to the genie decoder substantially, while the sliding-window and EKF variants trade some estimation accuracy for bounded latency, with a corresponding moderate increase in PLE.

For the $[[72,12,6]]$ BB-qLDPC code, the absolute PLE values and the spread across configurations are both larger than for the surface code. The uniform prior yields $p_L = 0.76$; for a code encoding $k = 12$ logical qubits, the saturated PLE ceiling is $1 - 2^{-k} \approx 0.9998$, so the uniform-prior decoder is already operating in a severely degraded regime. The genie decoder achieves $p_L = 0.12$, and the proposed algorithms yield $p_L = 0.18$ (Algorithm~1 offline), $p_L = 0.26$ (Algorithm~1 sliding window), and $p_L = 0.52$ (Algorithm~2), all substantially below the uniform baseline. The results here are for a specific muon samples and parameter configurations; the relative ordering of the algorithms may vary across different noise realizations or code geometries, and the absolute PLE values should be interpreted in that context.

\begin{figure}
    \centering
    \includegraphics[width=0.95\linewidth]{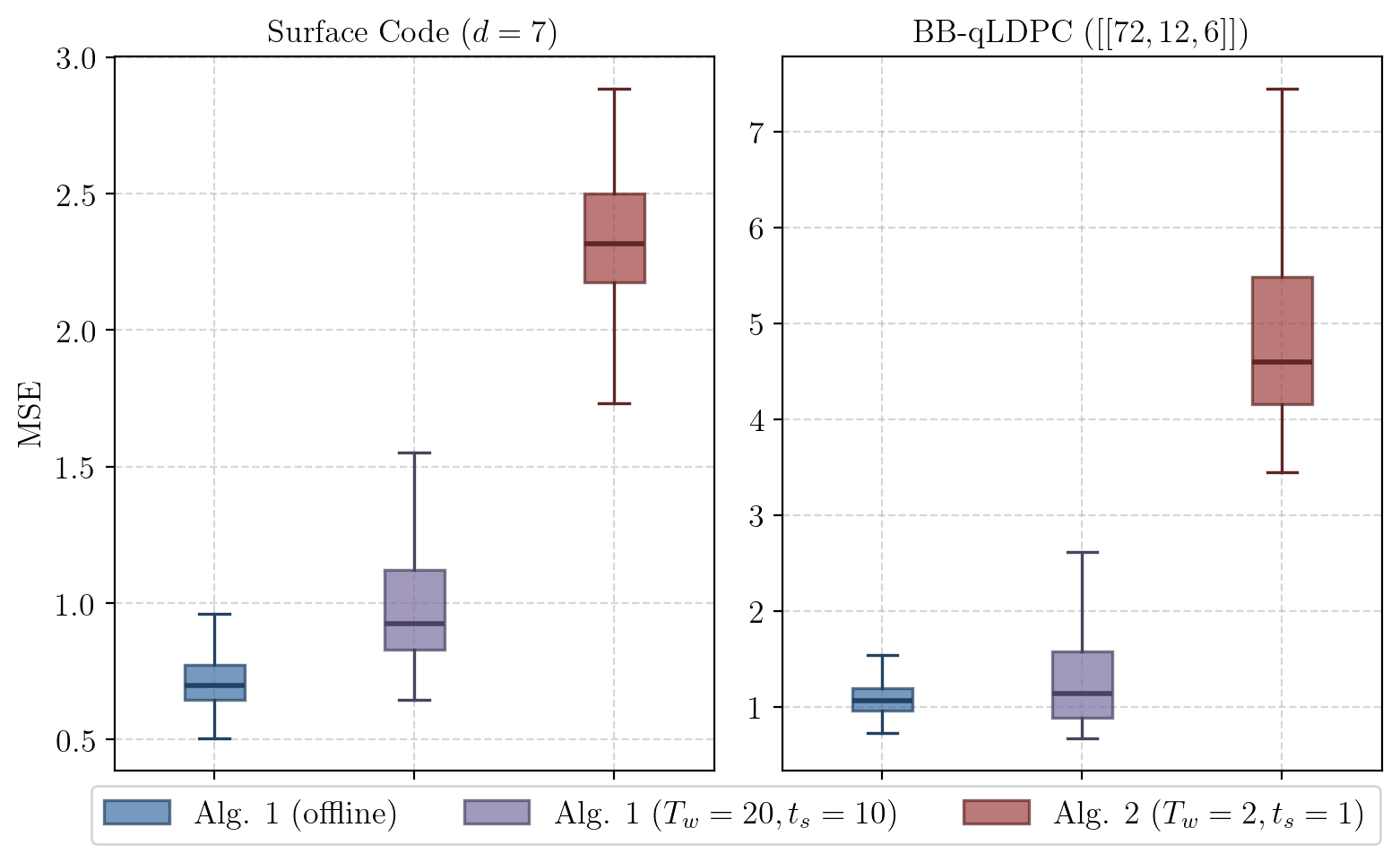}\vspace{-2mm}
    \caption{QP density estimation error of the proposed algorithms.}
    \label{fig:mse_boxplot}
\end{figure}

In the results presented, the PLE ordering across all configurations is consistent with the MSE ordering observed in Fig.~\ref{fig:mse_boxplot}: configurations with lower estimation error tend to achieve lower logical error rates. This empirical alignment suggests that QP estimation error is a meaningful driver of decoding performance for the configurations considered here. However, a formal monotone relationship between MSE and PLE has not been established; the correspondence observed here is across a specific set of algorithm configurations and two code instances, and need not hold in general. Nonetheless, the trend motivates treating estimation quality as a reasonable design objective when developing noise-aware decoders. Additional QP density estimation and decoding results provided in Appendix~\ref{apndx:additional_results}.

\subsection{Computational Complexity and Latency}
Table~\ref{table:complexity} summarizes the per-window computational complexity 
of the two algorithms, where $n$ is the number of qubits, $T_w$ is the window 
length, $K$ is the number of outer EM iterations, $I_{\rm bp}$ is the number of BP 
iterations, and $M$ is the number of inner gradient steps per EM iteration.

\begin{table}[h]
\caption{Computational complexity and Latency}
\centering
\renewcommand{\arraystretch}{1.6}
\begin{tabular}{lll}
\hline
\textbf{} & \textbf{Algorithm 1} & \textbf{Algorithm 2} \\
\hline
Total cost
  & $\mathcal{O}\bigl(K n T_w (I_{\mathrm{bp}} + M)\bigr)$
  & $\mathcal{O}\bigl(n T_w I_{\mathrm{bp}} + T_w n^3\bigr)$ \\
Memory
  & $\mathcal{O}(n T_w)$
  & $\mathcal{O}(n^2)$ \\
Latency
  & $\mathcal{O}\bigl(K n T_w ( I_{\mathrm{bp}} + M)\bigr)$
  & $\mathcal{O}\bigl(n T_w I_{\mathrm{bp}} + T_w n^3\bigr)$ \\
\hline
\end{tabular}
\label{table:complexity}
\end{table}

Alg.~1 performs $K$ outer iterations per window, each consisting of a BP pass costing $\mathcal{O}(n\,T_w\,I_{\rm bp})$ and $M$ gradient steps each costing $\mathcal{O}(n\,T_w)$, yielding a total of $\mathcal{O}(K\,n\,T_w\,(I_{\rm bp} + M))$ per window. The per-step gradient $\nabla_{Z_{t,i}}\mathcal{L}_{\rm err}$ is sparse, with the number of nonzero contributions per qubit determined by the column weight of the DEM check matrix, which is $\mathcal{O}(1)$ for LDPC codes such as surface codes and BB qLDPC codes. The per-step gradient $\nabla_{Z_{t_l:t_h}}\mathcal{L}_{\rm qp}$ requires two matrix-vector products with $A(\kappa)$ and $A(\kappa)^\top$ per time step; since $A(\kappa) = I - (sI + \kappa L)\Delta t$ inherits the sparsity of $L$, where each qubit is connected only to neighbors within radius $\rho$, the number of nonzeros per row is $\mathcal{O}(1)$ for fixed chip geometry, and each product costs $\mathcal{O}(n)$. Both gradient terms therefore scale as $\mathcal{O}(n\,T_w)$ per step, making Algorithm~1 well-suited to large sparse codes. The memory footprint is $\mathcal{O}(n\,T_w)$, as only the log QP density trajectory over the current window need be stored.

Alg.~2 replaces the inner gradient loop with a single forward EKF pass over the window. Each cycle requires one BP update at cost $\mathcal{O}(n\,I_{\rm bp})$ to form the pseudo-measurement, followed by EKF predict and update steps each costing $\mathcal{O}(n^3)$ due to the dense $n \times n$ covariance matrix operations. Summing over $T_w$ cycles gives a total cost of $\mathcal{O}(n\,T_w\,I_{\rm bp} + T_w\,n^3)$ per window, with the covariance update dominating for large $n$.\footnote{Even though the full covariance update costs $\mathcal{O}(n^3)$, structured approximations can reduce this substantially at the cost of approximation accuracy; we leave a systematic study of these trade-offs to future work.} The memory requirement is $\mathcal{O}(n^2)$ for storing $P_{t|t}$. Alg.~2 is therefore favorable in regimes where $n$ is moderate and the number of EM iterations $K\,M$ required by Alg.~1 is large, such as when the QP density varies rapidly across cycles.

\section{Conclusion and Future Directions}
We presented a joint decoding and sensing framework for stabilizer codes operating under radiation-induced correlated noise, in which the latent quasiparticle density field is explicitly modeled and inferred alongside Pauli error mechanisms. By casting syndrome-based decoding and QP tracking as a unified MAP inference problem and approximating its solution via an EM inspired alternating procedure, the decoder continuously adapts its error priors to the inferred spatiotemporal noise state. Two practical instantiations were developed: a gradient-based sliding-window algorithm that exploits the sparse temporal structure of the detector error model, and an extended Kalman filter variant that replaces the inner optimization loop with a recursive Bayesian estimator driven by pseudo-measurements derived from belief propagation posteriors. Both algorithms achieve this without any auxiliary sensing hardware, relying solely on the syndrome trajectory already available during standard QEC operation.

Simulations of a distance-7 rotated surface code and a $[[72, 12, 6]]$ bivariate bicycle qLDPC code under physically grounded QP dynamics generated from G4CMP show a substantial reduction in logical error probability relative to baseline decoders assuming a uniform noise prior, narrowing the gap toward a genie-aided decoder with perfect instantaneous knowledge of the QP trajectory. Beyond improved decoding performance, the inferred QP field yields physically meaningful diagnostics including spatial localization and temporal decay of radiation-induced events, directly linking logical-level behavior to chip-scale noise processes. This could have practical implications for device characterization, radiation shielding design, and fault monitoring. These results establish that integrating physical noise estimation into the decoding loop is beneficial for mitigating correlated noise in superconducting qubit systems.

Future work will address several extensions and limitations of the present study. The current formulation isolates radiation-induced noise; extending the model to jointly infer multiple latent noise sources and assessing robustness in mixed-noise conditions remains open. Co-design of qubit placement, code geometry, and measurement schedules to improve observability of correlated noise is a promising direction. We expect an active learning approach~\cite{zhao2021uncertainty} based on objective-driven uncertainty quantification~\cite{yoon2021quantifying} might be beneficial in efficiently navigating the huge co-design space based on multiple design objectives. Additional avenues include application to other stabilizer codes and fault-tolerant circuits, and extension to other forms of correlated noise such as cross-talk and control-induced correlations \cite{Gambetta2012, Preskill2018, BagewadiC2025}.

\bibliographystyle{IEEEtran}
\bibliography{bibfile}

\begin{thebibliography}{10}
\providecommand{\url}[1]{#1}
\csname url@samestyle\endcsname
\providecommand{\newblock}{\relax}
\providecommand{\bibinfo}[2]{#2}
\providecommand{\BIBentrySTDinterwordspacing}{\spaceskip=0pt\relax}
\providecommand{\BIBentryALTinterwordstretchfactor}{4}
\providecommand{\BIBentryALTinterwordspacing}{\spaceskip=\fontdimen2\font plus
\BIBentryALTinterwordstretchfactor\fontdimen3\font minus \fontdimen4\font\relax}
\providecommand{\BIBforeignlanguage}[2]{{%
\expandafter\ifx\csname l@#1\endcsname\relax
\typeout{** WARNING: IEEEtran.bst: No hyphenation pattern has been}%
\typeout{** loaded for the language `#1'. Using the pattern for}%
\typeout{** the default language instead.}%
\else
\language=\csname l@#1\endcsname
\fi
#2}}
\providecommand{\BIBdecl}{\relax}
\BIBdecl

\bibitem{vepsalainen2020impact}
A.~P. Veps{\"a}l{\"a}inen, A.~H. Karamlou, J.~L. Orrell, A.~S. Dogra, B.~Loer, F.~Vasconcelos, D.~K. Kim, A.~J. Melville, B.~M. Niedzielski, J.~L. Yoder, S.~Gustavsson, J.~A. Formaggio, B.~A. VanDevender, and W.~D. Oliver, ``Impact of ionizing radiation on superconducting qubit coherence,'' \emph{Nature}, vol. 584, no. 7822, pp. 551--556, 2020.

\bibitem{mcewen2022resolving}
M.~McEwen, L.~Faoro, K.~Arya, A.~Dunsworth, T.~Huang, S.~Kim, B.~Burkett, A.~Fowler, F.~Arute, J.~C. Bardin \emph{et~al.}, ``Resolving catastrophic error bursts from cosmic rays in large arrays of superconducting qubits,'' \emph{Nature Physics}, vol.~18, no.~1, pp. 107--111, 2022.

\bibitem{pattison2025hierarchical}
C.~A. Pattison, A.~Krishna, and J.~Preskill, ``Hierarchical memories: Simulating quantum {LDPC} codes with local gates,'' \emph{Quantum}, vol.~9, p. 1728, 2025.

\bibitem{trinca2022new}
C.~C. Trinca, C.~D. De~Albuquerque, R.~P. Junior, J.~C. Interlando, A.~A. De~Andrade, and R.~A. Watanabe, ``New quantum burst-error correcting codes from interleaving technique,'' in \emph{Proceedings of the 2022 IEEE Global Communications Conference (GLOBECOM)}, 2022, pp. 5243--5248.

\bibitem{elliott1963estimates}
E.~O. Elliott, ``Estimates of error rates for codes on burst-noise channels,'' \emph{The Bell System Technical Journal}, vol.~42, no.~5, pp. 1977--1997, 1963.

\bibitem{johnson2009burst}
S.~J. Johnson, ``Burst erasure correcting {LDPC} codes,'' \emph{IEEE Transactions on Communications}, vol.~57, no.~3, pp. 641--652, 2009.

\bibitem{andrews1997theory}
K.~Andrews, C.~Heegard, and D.~Kozen, ``A theory of interleavers,'' Cornell University, Tech. Rep., 1997.

\bibitem{hinton2002turbo}
D.~A. Hinton, ``Turbo coding in correlated fading channels,'' Ph.D. dissertation, Massachusetts Institute of Technology, 2002.

\bibitem{hou2001performance}
J.~Hou, P.~H. Siegel, and L.~B. Milstein, ``Performance analysis and code optimization of low density parity-check codes on {R}ayleigh fading channels,'' \emph{IEEE Journal on Selected Areas in Communications}, vol.~19, no.~5, pp. 924--934, 2001.

\bibitem{martinis2021saving}
J.~M. Martinis, ``Saving superconducting quantum processors from decay and correlated errors generated by gamma and cosmic rays,'' \emph{npj Quantum Information}, vol.~7, no.~1, p.~90, 2021.

\bibitem{yelton2024modeling}
E.~Yelton, C.~P. Larson, V.~Iaia, K.~Dodge, G.~La~Magna, P.~G. Baity, I.~V. Pechenezhskiy, R.~McDermott, N.~A. Kurinsky, G.~Catelani, and B.~L.~T. Plourde, ``Modeling phonon-mediated quasiparticle poisoning in superconducting qubit arrays,'' \emph{Physical Review B}, vol. 110, no.~2, p. 024519, 2024.

\bibitem{Gottesman1997}
D.~Gottesman, ``Stabilizer codes and quantum error correction,'' Ph.D. dissertation, California Institute of Technology, 1997.

\bibitem{Baity2026}
P.~G. Baity, A.~K. Nayak, L.~R. Varshney, N.~Jeon, B.-J. Yoon, P.~J. Love, and A.~Hoisie, ``Characterizing quantum error correction performance of radiation-induced errors,'' arXiv 2602.06202, 2026.

\bibitem{Fowler2012}
A.~G. Fowler, A.~M. Stephens, and P.~Groszkowski, ``{Surface codes: Towards practical large-scale quantum computation},'' \emph{Physical Review A}, vol.~86, no.~3, p. 032324, 2012.

\bibitem{gidney2021stim}
C.~Gidney, ``Stim: a fast stabilizer circuit simulator,'' \emph{Quantum}, vol.~5, p. 497, 2021.

\bibitem{bravyi2024high}
S.~Bravyi, A.~W. Cross, J.~M. Gambetta, D.~Maslov, P.~Rall, and T.~J. Yoder, ``High-threshold and low-overhead fault-tolerant quantum memory,'' \emph{Nature}, vol. 627, no. 8005, pp. 778--782, 2024.

\bibitem{Gallager1962LDPC}
R.~G. Gallager, ``Low-density parity-check codes,'' \emph{IRE Transactions on Information Theory}, vol.~8, no.~1, pp. 21--28, 1962.

\bibitem{Kschischang2001FactorGraphs}
F.~R. Kschischang, B.~J. Frey, and H.-A. Loeliger, ``Factor graphs and the sum-product algorithm,'' \emph{IEEE Transactions on Information Theory}, vol.~47, no.~2, pp. 498--519, 2001.

\bibitem{Forney2001FactorGraphs}
G.~D. Forney, Jr., ``The factor-graph approach to model-based signal processing,'' \emph{Proceedings of the IEEE}, vol.~89, no.~10, pp. 1563--1578, 2001.

\bibitem{MacKayNeal1997NearShannon}
D.~J.~C. MacKay and R.~M. Neal, ``Near {S}hannon limit performance of low-density parity-check codes,'' \emph{Electronics Letters}, vol.~32, no.~18, pp. 1645--1646, 1996.

\bibitem{PoulinChung2008SparseQuantumBP}
D.~Poulin and Y.~Chung, ``On the iterative decoding of sparse quantum codes,'' \emph{Quantum Information and Computation}, vol.~8, no.~10, pp. 987--1000, 2008.

\bibitem{FossorierLin1995OSD}
M.~P.~C. Fossorier and S.~Lin, ``Soft-decision decoding of linear block codes based on ordered statistics,'' \emph{IEEE Transactions on Information Theory}, vol.~41, no.~5, pp. 1379--1396, 1995.

\bibitem{Roffe2020DecodingLandscape}
J.~Roffe, D.~R. White, S.~Burton, and E.~T. Campbell, ``Decoding across the quantum low-density parity-check code landscape,'' \emph{Physical Review Research}, vol.~2, no.~4, p. 043423, 2020.

\bibitem{PanteleevKalachev2023DegenerateCodes}
P.~Panteleev and G.~Kalachev, ``Degenerate quantum codes and the quantum {H}amming bound,'' \emph{Quantum}, vol.~7, p. 1170, 2023.

\bibitem{yedidia2003understanding}
J.~S. Yedidia, W.~T. Freeman, and Y.~Weiss, ``Understanding belief propagation and its generalizations,'' in \emph{Exploring Artificial Intelligence in the New Millennium}.\hskip 1em plus 0.5em minus 0.4em\relax San Francisco, CA, USA: Morgan Kaufmann Publishers, Inc., 2003, pp. 239--269.

\bibitem{gong2024toward}
A.~Gong, S.~Cammerer, and J.~M. Renes, ``Toward low-latency iterative decoding of {QLDPC} codes under circuit-level noise,'' arXiv:2403.18901, 2024.

\bibitem{Kelsey2023}
M.~Kelsey, R.~Agnese, Y.~Alam, I.~A. Langroudy, E.~Azadbakht, D.~Brandt, R.~Bunker, B.~Cabrera, Y.-Y. Chang, H.~Coombes, R.~Cormier, M.~Diamond, E.~Edwards, E.~Figueroa-Feliciano, J.~Gao, P.~Harrington, Z.~Hong, M.~Hui, N.~Kurinsky, R.~Lawrence, B.~Loer, M.~Masten, E.~Michaud, E.~Michielin, J.~Miller, V.~Novati, N.~Oblath, J.~Orrell, W.~Perry, P.~Redl, T.~Reynolds, T.~Saab, B.~Sadoulet, K.~Serniak, J.~Singh, Z.~Speaks, C.~Stanford, J.~Stevens, J.~Strube, D.~Toback, J.~Ullom, B.~VanDevender, M.~Vissers, M.~Wilson, J.~Wilson, B.~Zatschler, and S.~Zatschler, ``{G4CMP}: Condensed matter physics simulation using the {Geant4} toolkit,'' \emph{Nuclear Instruments and Methods in Physics Research Section A: Accelerators, Spectrometers, Detectors and Associated Equipment}, vol. 1055, p. 168473, 2023.

\bibitem{Agostinelli2003}
S.~Agostinelli, J.~Allison, K.~Amako, J.~Apostolakis, H.~Araujo, P.~Arce, M.~Asai, D.~Axen, S.~Banerjee, G.~Barrand, F.~Behner, L.~Bellagamba, J.~Boudreau, L.~Broglia, A.~Brunengo, H.~Burkhardt, S.~Chauvie, J.~Chuma, R.~Chytracek, G.~Cooperman, G.~Cosmo, P.~Degtyarenko, A.~Dell'Acqua, G.~Depaola, D.~Dietrich, R.~Enami, A.~Feliciello, C.~Ferguson, H.~Fesefeldt, G.~Folger, F.~Foppiano, A.~Forti, S.~Garelli, S.~Giani, R.~Giannitrapani, D.~Gibin, J.~{Gómez Cadenas}, I.~González, G.~{Gracia Abril}, G.~Greeniaus, W.~Greiner, V.~Grichine, A.~Grossheim, S.~Guatelli, P.~Gumplinger, R.~Hamatsu, K.~Hashimoto, H.~Hasui, A.~Heikkinen, A.~Howard, V.~Ivanchenko, A.~Johnson, F.~Jones, J.~Kallenbach, N.~Kanaya, M.~Kawabata, Y.~Kawabata, M.~Kawaguti, S.~Kelner, P.~Kent, A.~Kimura, T.~Kodama, R.~Kokoulin, M.~Kossov, H.~Kurashige, E.~Lamanna, T.~Lampén, V.~Lara, V.~Lefebure, F.~Lei, M.~Liendl, W.~Lockman, F.~Longo, S.~Magni, M.~Maire, E.~Medernach, K.~Minamimoto, P.~{Mora de Freitas}, Y.~Morita, K.~Murakami, M.~Nagamatu,
  R.~Nartallo, P.~Nieminen, T.~Nishimura, K.~Ohtsubo, M.~Okamura, S.~O'Neale, Y.~Oohata, K.~Paech, J.~Perl, A.~Pfeiffer, M.~Pia, F.~Ranjard, A.~Rybin, S.~Sadilov, E.~{Di Salvo}, G.~Santin, T.~Sasaki, N.~Savvas, Y.~Sawada, S.~Scherer, S.~Sei, V.~Sirotenko, D.~Smith, N.~Starkov, H.~Stoecker, J.~Sulkimo, M.~Takahata, S.~Tanaka, E.~Tcherniaev, E.~{Safai Tehrani}, M.~Tropeano, P.~Truscott, H.~Uno, L.~Urban, P.~Urban, M.~Verderi, A.~Walkden, W.~Wander, H.~Weber, J.~Wellisch, T.~Wenaus, D.~Williams, D.~Wright, T.~Yamada, H.~Yoshida, and D.~Zschiesche, ``Geant4—a simulation toolkit,'' \emph{Nuclear Instruments and Methods in Physics Research Section A: Accelerators, Spectrometers, Detectors and Associated Equipment}, vol. 506, no.~3, pp. 250--303, 2003.

\bibitem{Allison2006}
J.~Allison, K.~Amako, J.~Apostolakis, H.~Araujo, P.~Arce~Dubois, M.~Asai, G.~Barrand, R.~Capra, S.~Chauvie, R.~Chytracek, G.~Cirrone, G.~Cooperman, G.~Cosmo, G.~Cuttone, G.~Daquino, M.~Donszelmann, M.~Dressel, G.~Folger, F.~Foppiano, J.~Generowicz, V.~Grichine, S.~Guatelli, P.~Gumplinger, A.~Heikkinen, I.~Hrivnacova, A.~Howard, S.~Incerti, V.~Ivanchenko, T.~Johnson, F.~Jones, T.~Koi, R.~Kokoulin, M.~Kossov, H.~Kurashige, V.~Lara, S.~Larsson, F.~Lei, O.~Link, F.~Longo, M.~Maire, A.~Mantero, B.~Mascialino, I.~McLaren, P.~Mendez~Lorenzo, K.~Minamimoto, K.~Murakami, P.~Nieminen, L.~Pandola, S.~Parlati, L.~Peralta, J.~Perl, A.~Pfeiffer, M.~Pia, A.~Ribon, P.~Rodrigues, G.~Russo, S.~Sadilov, G.~Santin, T.~Sasaki, D.~Smith, N.~Starkov, S.~Tanaka, E.~Tcherniaev, B.~Tome, A.~Trindade, P.~Truscott, L.~Urban, M.~Verderi, A.~Walkden, J.~Wellisch, D.~Williams, D.~Wright, and H.~Yoshida, ``Geant4 developments and applications,'' \emph{IEEE Transactions on Nuclear Science}, vol.~53, no.~1, pp. 270--278, 2006.

\bibitem{Allison2016}
J.~Alison, K.~Amako, J.~Apostolakis, P.~Arce, M.~Asai, T.~Aso, E.~Bagli, A.~Bagulya, S.~Banerjee, G.~Barrand, B.~Beck, A.~Bogdanov, D.~Brandt, J.~Brown, H.~Burkhardt, P.~Canal, D.~Cano-Ott, S.~Chauvie, K.~Cho, G.~Cirrone, G.~Cooperman, M.~Cortés-Giraldo, G.~Cosmo, G.~Cuttone, G.~Depaola, L.~Desorgher, X.~Dong, A.~Dotti, V.~Elvira, G.~Folger, Z.~Francis, A.~Galoyan, L.~Garnier, M.~Gayer, K.~Genser, V.~Grichine, S.~Guatelli, P.~Guèye, P.~Gumplinger, A.~Howard, I.~Hřivnáčová, S.~Hwang, S.~Incerti, A.~Ivanchenko, V.~Ivanchenko, F.~Jones, S.~Jun, P.~Kaitaniemi, N.~Karakatsanis, M.~Karamitros, M.~Kelsey, A.~Kimura, T.~Koi, H.~Kurashige, A.~Lechner, S.~Lee, F.~Longo, M.~Maire, D.~Mancusi, A.~Mantero, E.~Mendoza, B.~Morgan, K.~Murakami, T.~Nikitina, L.~Pandola, P.~Paprocki, J.~Perl, I.~Petrović, M.~Pia, W.~Pokorski, J.~Quesada, M.~Raine, M.~Reis, A.~Ribon, A.~Ristić~Fira, F.~Romano, G.~Russo, G.~Santin, T.~Sasaki, D.~Sawkey, J.~Shin, I.~Strakovsky, A.~Taborda, S.~Tanaka, B.~Tomé, T.~Toshito, H.~Tran,
  P.~Truscott, L.~Urban, V.~Uzhinsky, J.~Verbeke, M.~Verderi, B.~Wendt, H.~Wenzel, D.~Wright, D.~Wright, T.~Yamashita, J.~Yarba, and H.~Yoshida, ``Recent developments in {Geant4},'' \emph{Nuclear Instruments and Methods in Physics Research Section A: Accelerators, Spectrometers, Detectors and Associated Equipment}, vol. 835, pp. 186--225, 2016.

\bibitem{google2023suppressing}
{Google Quantum AI}, ``Suppressing quantum errors by scaling a surface code logical qubit,'' \emph{Nature}, vol. 614, no. 7949, pp. 676--681, 2023.

\bibitem{akiba2019optuna}
T.~Akiba, S.~Sano, T.~Yanase, T.~Ohta, and M.~Koyama, ``Optuna: A next-generation hyperparameter optimization framework,'' in \emph{Proceedings of the 25th ACM SIGKDD International Conference on Knowledge Discovery \& Data Mining}, 2019, pp. 2623--2631.

\bibitem{zhao2021uncertainty}
G.~Zhao, E.~Dougherty, B.-J. Yoon, F.~Alexander, and X.~Qian, ``Uncertainty-aware active learning for optimal {B}ayesian classifier,'' in \emph{Proceedings of the International Conference on Learning Representations (ICLR)}, 2021.

\bibitem{yoon2021quantifying}
B.-J. Yoon, X.~Qian, and E.~R. Dougherty, ``Quantifying the multi-objective cost of uncertainty,'' \emph{IEEE Access}, vol.~9, pp. 80\,351--80\,359, 2021.

\bibitem{Gambetta2012}
J.~M. Gambetta, A.~D. C{\'{o}}rcoles, S.~T. Merkel, B.~R. Johnson, J.~A. Smolin, J.~M. Chow, C.~A. Ryan, C.~Rigetti, S.~Poletto, T.~A. Ohki, M.~B. Ketchen, and M.~Steffen, ``{Characterization of addressability by simultaneous randomized benchmarking},'' \emph{{Physical Review Letters}}, vol. 109, no.~24, p. 240504, 2012.

\bibitem{Preskill2018}
J.~Preskill, ``{Quantum computing in the NISQ era and beyond},'' \emph{{Quantum}}, vol.~2, p.~79, 2018.

\bibitem{BagewadiC2025}
S.~Bagewadi and A.~Chatterjee, ``Effect of correlated errors on quantum memory,'' \emph{Physical Review A}, vol. 112, p. 022422, Aug. 2025.

\end{thebibliography}

\appendix

\subsection{Parity Check Matrix: Spatiotemporal Coupling}
\label{apndx:pcm}

Under circuit-level noise, error mechanisms can be classified by the rounds whose detectors they affect: \emph{intra-round} error mechanisms $\widetilde{E}^{(0)}_t$ whose detector support lies entirely within round $t$, and \emph{spanning} error mechanisms $\widetilde{E}^{(1)}_t$ that flip detectors in both rounds $t$ and $t{+}1$. The syndrome $D_t$ depends on at most three subblocks of the global parity check matrix,
\begin{equation}
    D_{t} = \bigl[H^{(t-1)}_2 \;\big|\; H^{(t)}_0 \;\big|\; H^{(t)}_1\bigr]
    \bigl[\widetilde{E}^{(1)}_{t-1} \;\big|\; \widetilde{E}^{(0)}_{t} \;\big|\; 
    \widetilde{E}^{(1)}_{t}\bigr]^{\!\top},
\end{equation}
where $H^{(t)}_0$, $H^{(t)}_1$, and $H^{(t)}_2$ are the submatrices coupling round-$t$ detectors to intra-round error mechanisms, forward-spanning error mechanisms originating in round $t$, and forward-spanning error mechanisms originating in round $t-1$, respectively. Stacking over all rounds yields the block structure shown below, with each block row overlapping its neighbor via the shared subblock $H^{(t+1)}_2$.
{
\setlength{\arraycolsep}{2pt}
\[
\begin{bmatrix}
D_1 \\ D_2 \\ D_3 \\ \vdots
\end{bmatrix} = 
\begin{bmatrix}
H_0^{(1)} & H_1^{(1)} & & & & & \\
& H_2^{(1)} & H_0^{(2)} & H_1^{(2)} & & & \\
& & & H_2^{(2)} & H_0^{(3)} & H_1^{(3)} & \\
& & & & &  & \ddots \\
\end{bmatrix}
\begin{bmatrix}
\widetilde{E}_1^{(1)} \\
\widetilde{E}_1^{(2)} \\
\widetilde{E}_2^{(1)} \\
\widetilde{E}_2^{(2)} \\
\widetilde{E}_3^{(1)} \\
\widetilde{E}_3^{(2)} \\
\vdots
\end{bmatrix}
\]
}

The sliding window decoding \cite{gong2024toward} advances by one stride after updating the boundary syndrome
\begin{equation}
D_2 := D_2 \oplus H_2^{(1)}\,\widetilde{E}^{*\,(2)}_1
\end{equation}
to absorb the committed spanning contribution, and the next window solves an identical $3\times 4$ block system with $(D_2, D_3, D_4)$ on the right-hand side. Repeating with the general update $D_{t+1} := D_{t+1} \oplus H_2^{(t)}\widetilde{E}^{*\,(2)}_t$ at each stride, the committed partial equations help in satisfying the global parity-check equation $H \tilde{E}_{1:T} = D_{1:T}$.

\subsection{Variational Decomposition}

\label{apndx:variational_decomposition}

We begin by introducing an arbitrary variational distribution $q(E_{1:T})$ over the Pauli fault trajectory and applying the standard evidence lower bound identity to decompose the log joint. Multiplying and dividing by $q(E_{1:T})$ inside the expectation and rearranging yields
\begin{align}
\log& ~p(X_{1:T}, \theta, D_{1:T}) \\
& = \mathbb{E}_{q(E_{1:T})}
\!\left[
\log
\frac{
p(X_{1:T}, \theta, D_{1:T}, E_{1:T})
}{
p(E_{1:T} \mid X_{1:T}, \theta, D_{1:T})
}
\right] \nonumber \\
&= \log~p(X_{1:T}, \theta)
+ \mathbb{E}_{q(E_{1:T})}
\!\left[
\log p(E_{1:T} \mid X_{1:T})
\right] \nonumber \\
& \quad + H(q(E_{1:T})) 
+ \mathrm{KL}\!\left(
q(E_{1:T})
\,\|\, p(E_{1:T} \mid D_{1:T}, X_{1:T})
\right).
\end{align}

The KL divergence term vanishes if and only if $q(E_{1:T}) = p(E_{1:T} \mid D_{1:T}, X_{1:T})$, which would constitute an ideal decoder. However, exact posterior inference over the full Pauli fault trajectory is computationally intractable due to the exponential state space of Pauli faults. We therefore introduce two auxiliary distributions: $p_b(\tilde{E}_{1:T} \mid D_{1:T}, X_{1:T})$, the BP posterior over error mechanisms (Bethe distribution), and $p_m(\tilde{E}_{1:T} \mid X_{1:T})$, the marginal prior over error mechanisms induced by the noise model. Multiplying and dividing by these quantities inside the expectation and rearranging yields
\begin{align*}
&\log~p(X_{1:T}, \theta, D_{1:T})\\
&= \mathbb{E}_q \!\left[
\log \frac{p(X_{1:T}, \theta, D_{1:T}, E_{1:T}, \tilde{E}_{1:T})}{p(E_{1:T}, \tilde{E}_{1:T} \mid X_{1:T}, \theta, D_{1:T})}\right] \\
&= \mathbb{E}_{q} \!\left[
\log\frac{p(X_{1:T}, \theta, D_{1:T}, E_{1:T}, \tilde{E}_{1:T})}{p(E_{1:T}, \tilde{E}_{1:T} \mid X_{1:T}, \theta, D_{1:T})}
\frac{q(E_{1:T}, \tilde{E}_{1:T})}{q(E_{1:T}, \tilde{E}_{1:T})} \right. \\
&\qquad\qquad\left.
\frac{p_b(\tilde{E}_{1:T} \mid D_{1:T}, X_{1:T})}
     {p_b(\tilde{E}_{1:T} \mid D_{1:T}, X_{1:T})}
\frac{p_m(\tilde{E}_{1:T} \mid X_{1:T})}
     {p_m(\tilde{E}_{1:T} \mid X_{1:T})}
\right] \\
&= \mathbb{E}_{q} \!\left[
\log \Bigg(p(X_{1:T}, \theta) \frac{p_m(\tilde{E}_{1:T} \mid X_{1:T}) q(\tilde{E}_{1:T})}{q(\tilde{E}_{1:T}) p_b(\tilde{E}_{1:T} \mid D_{1:T}, X_{1:T})}\Bigg)
\right. \\
&\qquad
\left. \Bigg(\frac{p(E_{1:T}, \tilde{E}_{1:T}, D_{1:T} \mid X_{1:T})}{p(E_{1:T}, \tilde{E}_{1:T} \mid X_{1:T}, D_{1:T})} \frac{q(E_{1:T} \mid \tilde{E}_{1:T})} {q(E_{1:T} \mid \tilde{E}_{1:T})} \right. \\
& \qquad \qquad \left.  \frac{p_b(\tilde{E}_{1:T} \mid D_{1:T}, X_{1:T})}{p_m(\tilde{E}_{1:T} \mid X_{1:T})}\Bigg)\right].
\end{align*}
Since the mapping from fault trajectory to detection events is deterministic, $p(D_{1:T} \mid E_{1:T}, \tilde{E}_{1:T}, X_{1:T}) = 1$ for any fault trajectory consistent with the observed syndrome, which simplifies the remaining expectation to yield
\begin{align*}
&= \log p(X_{1:T}, \theta) - \mathrm{KL}\!\left(
q(\tilde{E}_{1:T})\,\|\, p_m(\tilde{E}_{1:T} \mid X_{1:T})\right) \\
&+ \mathrm{KL}\!\left(q(\tilde{E}_{1:T})\,\|\, p_b(\tilde{E}_{1:T} \mid D_{1:T}, X_{1:T})\right) \\
&+ \mathbb{E}_q \!\left[\log\frac{p(E_{1:T}, \tilde{E}_{1:T} \mid X_{1:T})
\, \overbrace{p(D_{1:T} \mid E_{1:T}, \tilde{E}_{1:T}, X_{1:T})}^{\text{deterministic}}}{p(E_{1:T}, \tilde{E}_{1:T} \mid X_{1:T}, D_{1:T})}\right. \\
&\qquad\qquad\left.
\frac{q(E_{1:T} \mid \tilde{E}_{1:T})}
     {q(E_{1:T} \mid \tilde{E}_{1:T})}
\frac{p_b(\tilde{E}_{1:T} \mid D_{1:T}, X_{1:T})}
     {p_m(\tilde{E}_{1:T} \mid X_{1:T})}
\right]
\end{align*}

\begin{align*}
&= \log p(X_{1:T}, \theta)
- \mathrm{KL}\!\left(
q(\tilde{E}_{1:T})
\,\|\, p_m(\tilde{E}_{1:T} \mid X_{1:T})
\right) \\
&\quad
+ \mathrm{KL}\!\left(
q(\tilde{E}_{1:T})
\,\|\, p_b(\tilde{E}_{1:T} \mid D_{1:T}, X_{1:T})
\right) \\
&\quad
+ \mathbb{E}_q \!\left[
\log
\frac{
p(E_{1:T} \mid \tilde{E}_{1:T}, X_{1:T})
\, p(\tilde{E}_{1:T} \mid X_{1:T})
}{
p(E_{1:T} \mid \tilde{E}_{1:T}, X_{1:T}, D_{1:T})
\, p(\tilde{E}_{1:T} \mid X_{1:T}, D_{1:T})
} \right.\\
& \qquad \qquad \left.
\frac{q(E_{1:T} \mid \tilde{E}_{1:T})
\, p_b(\tilde{E}_{1:T} \mid D_{1:T}, X_{1:T})}{q(E_{1:T} \mid \tilde{E}_{1:T})
\, p_m(\tilde{E}_{1:T} \mid X_{1:T})}
\right].
\end{align*}

Finally, applying the chain rule of probability and the Markov structure of the graphical model to factor the joint over $(E_{1:T}, \tilde{E}_{1:T})$, the last expectation separates into contributions from $q(E_{1:T} \mid \tilde{E}_{1:T})$ and $q(\tilde{E}_{1:T})$, yielding the final decomposition

\begin{align*}
&= \log p(X_{1:T}, \theta)
- \mathrm{KL}\!\left(q(\tilde{E}_{1:T})\,\|\, p_m(\tilde{E}_{1:T} \mid X_{1:T})\right) \\
&\quad + \mathrm{KL}\!\left(q(\tilde{E}_{1:T})\,\|\, p_b(\tilde{E}_{1:T} \mid D_{1:T}, X_{1:T})\right) \\
&\quad
+ \underbrace{\mathbb{E}_q \!\left[\log
\frac{p(E_{1:T} \mid \tilde{E}_{1:T}, X_{1:T})}{p(E_{1:T} \mid \tilde{E}_{1:T}, X_{1:T}, D_{1:T})} \right]
}_{\text{\hspace{4cm} degeneracy residual}} 
\\
&\quad + \underbrace{
\mathbb{E}_q \!\left[\log\frac{p(\tilde{E}_{1:T} \mid X_{1:T})}{p_m(\tilde{E}_{1:T} \mid X_{1:T})}\right]}_{\hspace{1.5cm}{\text{mean-field approximation error}}}
\\
&\quad + \underbrace{\mathbb{E}_q \!\left[\log\frac{p_b(\tilde{E}_{1:T} \mid D_{1:T}, X_{1:T})}{p(\tilde{E}_{1:T} \mid D_{1:T}, X_{1:T})}
\right]}_{\hspace{4cm}{\text{algorithmic error}}}.
\end{align*}

This final form reveals the structure: the first KL divergence term penalizes $q(\tilde{E}_{1:T})$ for deviating from the noise-model prior $p_m$, the second KL divergence term incentivizes $q(\tilde{E}_{1:T})$ to match the BP posterior $p_b$ over error mechanisms, and the remaining expectation captures the residual approximation error 
introduced by degeneracy, mean-field approximation, and belief propagation decoding algorithm.

\subsection{Gradient Expressions}

\subsubsection{Expression for $\nabla_{Z_t} \mathcal{L}_{\rm qp}$}

The QP dynamics loss is a weighted least-squares penalty in log-space,
\begin{equation}
    \mc{L}_{\rm qp}(X_{1:T}, \kappa) = -\frac{1}{2}\sum_{t=1}^{T} 
    \|Z_t - \log(A(\kappa)\,e^{Z_{t-1}})\|^2_{Q^{-1}}.
\end{equation}
Differentiating with respect to $Z_t$ requires accounting for two contributions: 
$Z_t$ appears explicitly as the current state, and also implicitly as $Z_{t-1}$ 
in the next time step. Defining the residual
\begin{equation}
    \varepsilon_t := Z_t - \log\!\left(A(\kappa)\,e^{Z_{t-1}}\right),
\end{equation}
the gradient is
\begin{equation}
    \nabla_{Z_t} \mathcal{L}_{\rm qp} = Q^{-1}\varepsilon_t 
    - e^{Z_t} \odot A(\kappa)^\top\!\left(Q^{-1}\varepsilon_{t+1} 
    \oslash \left(A(\kappa)\,e^{Z_t}\right)\right),
\end{equation}
where $\odot$ and $\oslash$ denote element-wise multiplication and division 
respectively. The first term is the direct gradient from the $t$th residual, 
and the second term is the back-propagated contribution from the $(t+1)$th 
residual through the nonlinear diffusion map.

\subsubsection{Expression for $\nabla_{Z_{t,i}} \mathcal{L}_{\rm err}$}

The error likelihood gradient couples the QP density at qubit $i$, time $t$ to 
all error mechanisms whose Tanner graph support includes a Pauli fault on qubit 
$i$ at time $t$. Denoting the BP posterior and prior probabilities of error mechanism $\tilde{\mc{E}}_{t',j}$ as $q_{t',j} := q\!\left(\tilde{E}_{t',j} {=} 1 \mid D_{1:T},\, X_{1:T}^{(k)}\right)$, and $p_{t',j} := p_m\!\left(\tilde{E}_{t',j} {=} 1 \mid X_{1:T}\right)$, the gradient is obtained by applying the chain rule through the change of variables 
$Z_{t,i} = \log X_{t,i}$,
\begin{equation}
\frac{\partial \mathcal{L}_{\text{err}}}{\partial Z_{t,i}} = \hspace{-5mm}
\sum_{\substack{(t',j,\tau):\\ \Phi(E_{t,\tau,i}) \in \tilde{\mathcal{E}}_{t',j}}}
\left[ \frac{q_{t',j}}{p_{t',j}} - \frac{1 - q_{t',j}}{1 - p_{t',j}} \right] 
X_{t,i}\, \frac{\partial p(E_{t,\tau,i} \mid X_{t,i})}{\partial X_{t,i}}.
\end{equation}
The bracketed term is the log-likelihood residual for error mechanism 
$\tilde{\mc{E}}_{t',j}$: it is positive when the BP posterior assigns higher 
probability to the error mechanism than the current QP density predicts, driving 
$Z_{t,i}$ upward, and negative otherwise. The sum extends over all fault types $\tau$ and error mechanisms $(t', j)$ that have qubit $i$ at time $t$ 
in their support, reflecting the temporal coupling in the DEM Tanner graph 
discussed in Section~\ref{subsec:dem}.

\subsection{Extended Kalman Filter Equations}
\label{apndx:kalman_equations}

We summarize the EKF recursion for the log-space state $Z_t = \log X_t \in \mathbb{R}^J$,
where $J$ is the number of sites where QP density is to be tracked. The filter alternates between a predict step, which 
propagates the state estimate forward through the nonlinear diffusion map, and an update step, which incorporates the pseudo-measurement $Y_t$ derived from the BP posteriors.

\paragraph{Predict Step}
Given the posterior estimate $\hat{Z}_{t-1|t-1}$ and covariance $P_{t-1|t-1}$ from the 
previous cycle, the prior estimate at time $t$ is obtained by applying the diffusion map 
in the original (non-log) domain and transforming back to log-space,
\begin{equation}
    \hat{Z}_{t|t-1} = \log\!\left(A(\kappa)\, \exp(\hat{Z}_{t-1|t-1})\right),
\end{equation}
where $\exp(\cdot)$ and $\log(\cdot)$ are applied elementwise. Since this map is 
nonlinear, we propagate the covariance using the first-order Jacobian $F_t$, whose 
$(i,j)$th entry is
\begin{equation}
    [F_t]_{ij} = \frac{[A(\kappa)]_{ij}\, e^{\hat{Z}_{t-1|t-1,j}}}
    {\left[A(\kappa)\exp(\hat{Z}_{t-1|t-1})\right]_i},
\end{equation}
yielding the predicted covariance
\begin{equation}
    P_{t|t-1} = F_t\, P_{t-1|t-1}\, F_t^T + Q.
\end{equation}

\paragraph{Update Step}
Given the pseudo-measurement $Y_t$ and the Gaussian measurement model 
$\log Y_t \mid Z_t \sim \mc{N}(Z_t, R)$, the innovation, innovation covariance, 
and Kalman gain are
\begin{align}
    r_t &= \log Y_t - \hat{Z}_{t|t-1}, \\
    S_t &= P_{t|t-1} + R, \\
    \mc{K}_t &= P_{t|t-1}\, S_t^{-1}.
\end{align}
The posterior mean and covariance are then
\begin{align}
    \hat{Z}_{t|t} &= \hat{Z}_{t|t-1} + \mc{K}_t\, r_t, \\
    P_{t|t} &= (\mathbb{I}_J - \mc{K}_t)\, P_{t|t-1}.
\end{align}
Here $Q = \sigma_Q^2\,\mathbbm{I}_J$ and $R = \sigma_R^2\,\mathbbm{I}_J$ are isotropic 
process and measurement noise covariances respectively, and $\mathbb{I}_J$ denotes the 
$J \times J$ identity matrix.

\subsection{Parameter Selection}

\label{apndx:bayes_param_select}
We tune the parameters for each code layout using Bayesian optimization using Optuna's tree-structured Parzen estimator (TPE) \cite{akiba2019optuna}. Alg.~1 with full-horizon is treated as the representative case, and tuned parameters are used for all other algorithm choices: the search space covers: diffusion coefficient $\kappa \in [0.1, 0.7]$, QP trapping rate $s \in [0, 0.5]$, the process noise variance $\sigma_Q \in [0.01, 10.0]$.

Each optimization run has 30 sequential rounds, each evaluating a batch of 5 parameter configurations. For each configuration, the joint decoding and QP estimation is run on 10 muon samples (which does not include both sample 53 and 58), and the average maximum MSE ($\max_{t}\frac{1}{J} \sum_{t,i} (Z^*_{t,i} - \log X_{t,i})^2$) over these samples is reported to the TPE surrogate, yielding $10 \times 5 = 50$ total trials per run. The first 3 rounds are for pure random exploration; the TPE surrogate activates from round 4 onward. The best parameters are chosen as those that minimize the average of maximum MSE, where MSE is computed for each cycle, maximum is across all cycles (time), and average is across all muon samples and are reported in Table~\ref{tab:bayes_best_params}.

\begin{table}[h]
\centering
\caption{Bayes-Optimal Parameters for Algorithm~1}\vspace{-3mm}
\label{tab:bayes_best_params}
\begin{tabular}{lcccc}
\toprule
Code & Avg.\ Max MSE & $\kappa$ & $\sigma^2_Q$ & $s$ \\
\midrule
Surface code ($d=7$)       & 1.7871 & 0.6367 & 2.2889 & 0.0498 \\
BB-qLDPC $[\![72,12,6]\!]$ & 2.1382 & 0.1433 & 1.3735 & 0.0012 \\
\bottomrule
\end{tabular}
\end{table}

\subsection{Experiments}

\label{apndx:additional_results}
\subsubsection{Highest Impact Muon Sample Selection}

To characterize variability across radiation events, we evaluate PLE ratio $p_L^{(\mathrm{uniform})} / p_L^{(\mathrm{genie})}$ for each of the 64 muon samples in the dataset. We choose Algorithm~\ref{algo:decode_sense_sliding_window} under full-horizon decoding with $T_w = T =$ 50 \textmu s.

For distance-7 surface code, the mean genie PLE is $4.202 \times 10^{-2}$ and the mean uniform PLE is $5.342 \times 10^{-2}$, which is $1.27\times$ improvement from perfect knowledge of QP densities. The mean PLE ratio of $1.167$. The distribution is right-skewed: most samples show modest degradation under the uniform prior, while a small subset of high-impact strikes produce substantially larger ratios.
Sample 58, with the highest PLE ratio of $3.424$, is selected as the worst-case muon sample.

For \([[72,12,6]]\) BB-qLDPC code, the mean genie PLE is $2.180 \times 10^{-2}$ and the mean uniform PLE is $5.815 \times 10^{-2}$, which is $2.67\times$ improvement from perfect prior knowledge. The mean PLE ratio is $2.181$. Sample 53, with the highest per-sample ratio of $4.763$, is selected as the representative muon sample.

The $[[72,12,6]]$ BB-qLDPC code exhibits consistently higher PLE ratios than the surface code. Two factors potentially contribute:
First, the BB-qLDPC layout occupies a larger physical footprint (cf.\ Figs.~\ref{fig:code_layout_xqp_surface_xqp58} and~\ref{fig:code_layout_xqp_bb_qldpc_xqp53}),
so a muon strike affects a larger fraction of encoded qubits, producing
stronger spatial correlations that a uniform prior cannot capture.
Second, the higher (effective) code rate ($k/N = 1/12$ vs.\ $1/97$) means fewer physical qubits per logical qubit, increasing sensitivity to
prior mismatch under correlated noise. This suggests that high-rate qLDPC codes could benefit most from accurate QP density estimation.

\begin{figure}
    \centering
    \includegraphics[width=0.99\linewidth]{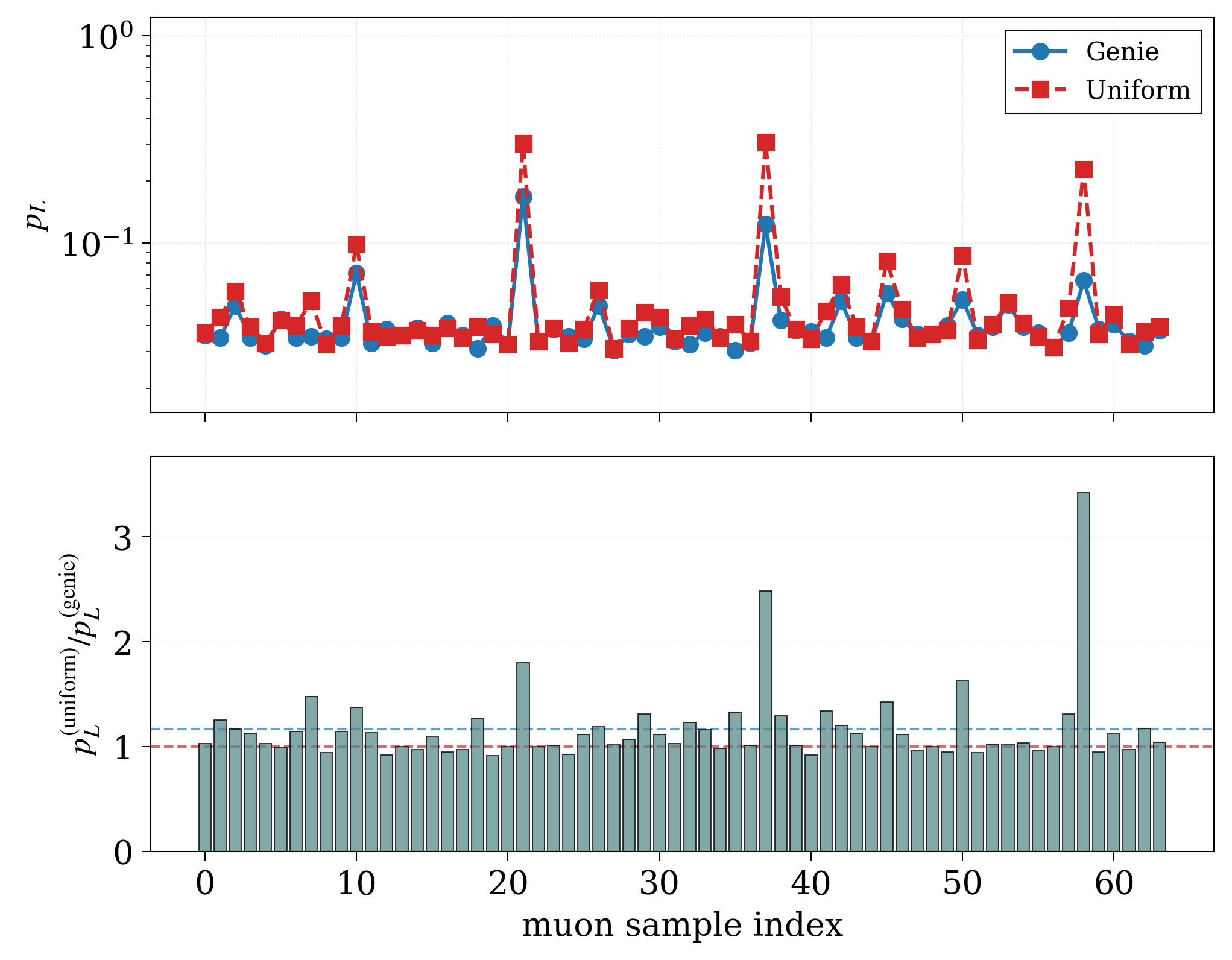}\vspace{-3mm}
    \caption{Ratio of PLE with uniform prior to PLE with genie prior for $[[72,12,6]]$ surface code layout shown in Figure~\ref{fig:code_layout_xqp_surface_xqp58} across all muon samples.}
    \label{fig:ple_genie_vs_uniform_surface}
\end{figure}

\begin{figure}
    \centering
    \includegraphics[width=0.99\linewidth]{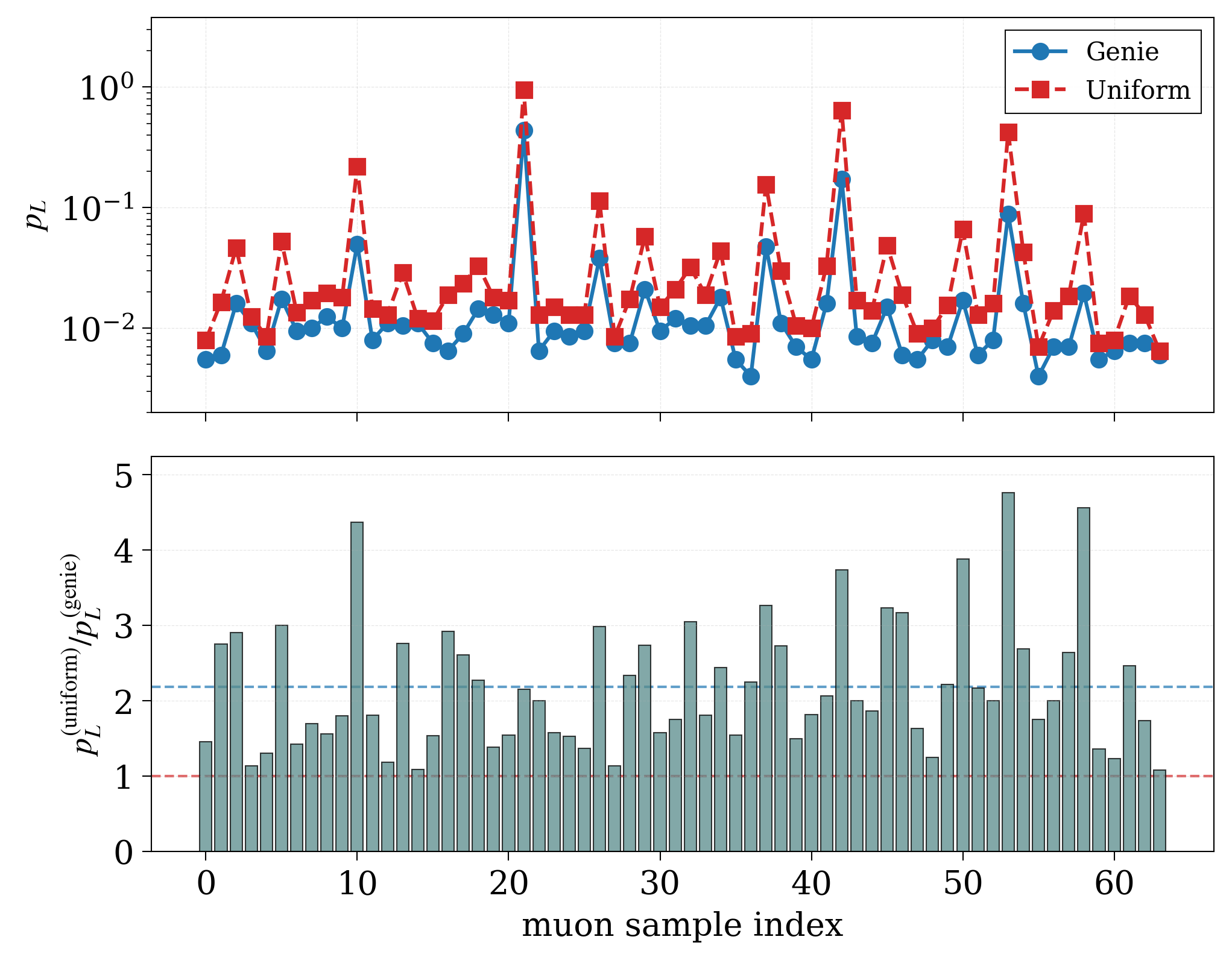}\vspace{-3mm}
    \caption{Ratio of PLE with uniform prior to PLE with genie prior for $[[72,12,6]]$ BB-qLDPC code layout shown in Fig.~\ref{fig:code_layout_xqp_bb_qldpc_xqp53} across all muon samples.}
    \label{fig:ple_genie_vs_uniform_bbqldpc}
\end{figure}

\subsubsection{QP Density Estimation}

\begin{figure}
    \centering
    \includegraphics[width=0.99\linewidth]{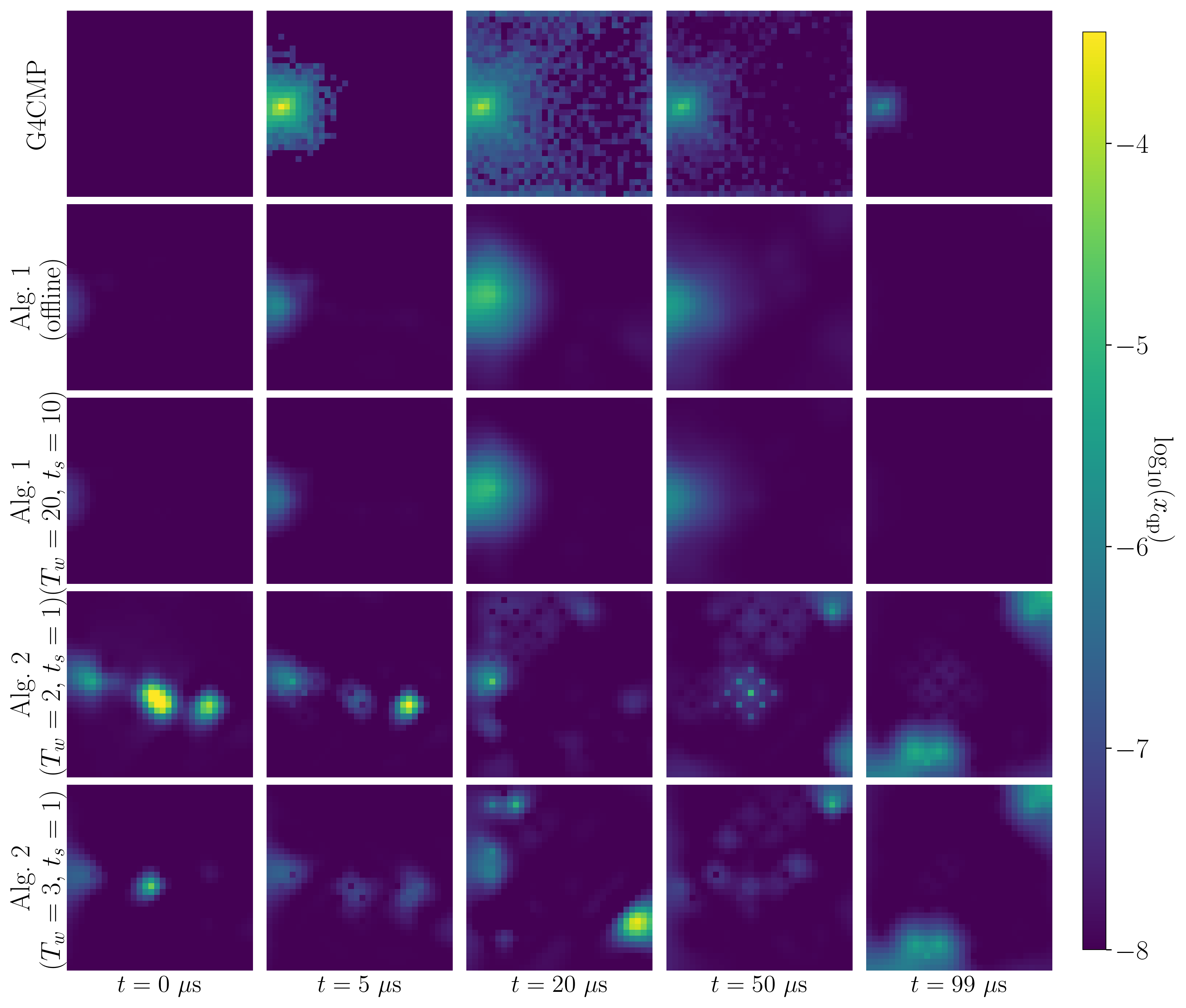}\vspace{-3mm}
    \caption{QP density estimates at time slices $t=$0 \textmu s, 5 \textmu s, 20 \textmu s, 50 \textmu s, and 99 \textmu s using Alg.~\ref{algo:decode_sense_sliding_window} (full-horizon and sliding window), and Alg.~\ref{algo:kalman} (online), with muon sample 58 and distance 7 surface code.}
    \label{fig:xqp58_est_surface_d7}
\end{figure}

\begin{figure}
    \centering
    \includegraphics[width=0.99\linewidth]{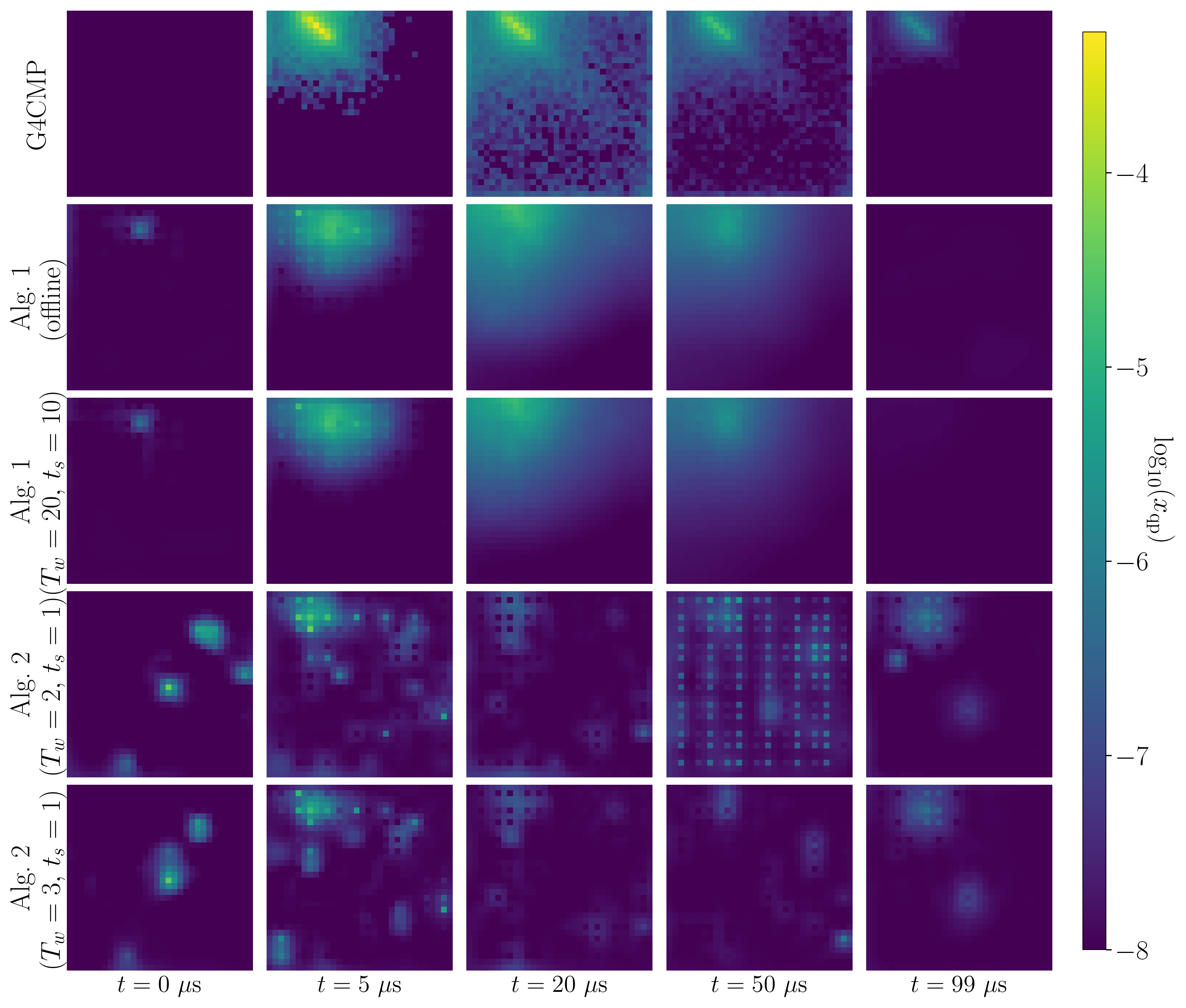}\vspace{-3mm}
    \caption{QP density estimates at time slices $t=0$ \textmu s, 5 \textmu s, 20 \textmu s, 50 \textmu s, and 99 \textmu s using Alg.~\ref{algo:decode_sense_sliding_window} (full-horizon and sliding window), and Alg.~\ref{algo:kalman} (online), with muon sample 53 and $[[72,12,6]]$ BB-qLDPC code.}
    \label{fig:xqp53_est_bbqldpc_n72}
\end{figure}

Figs.~\ref{fig:xqp58_est_surface_d7} and~\ref{fig:xqp53_est_bbqldpc_n72} display QP density estimates at 5 time slices ($t = 0, 5, 20, 50, 99\,\mu$s), interpolated to non-qubit locations via spline RBF in log-space.

Alg.~\ref{algo:decode_sense_sliding_window} (offline) accurately recovers the burst peak and spatial extent because the full-horizon gradient couples all syndrome cycles simultaneously, allowing backward information from post-burst decay to sharpen the burst-phase estimate. The sliding-window variant ($T_w = 20$, $t_s = 10$) mildly underestimates the burst peak since the leading window edge has not yet accumulated sufficient syndrome history to fully resolve the impact.

Alg.~\ref{algo:kalman} ($T_w = 2$, $t_s = 1$) produces spurious high-density artifacts at isolated qubit sites at $t = 0\,\mu$s, attributable to false alarms: the minimal window length provides insufficient syndrome history to suppress noise-driven pseudo-measurement fluctuations, and the log-normal approximation amplifies these fluctuations. Extending to $T_w = 3$ substantially suppresses these artifacts by averaging the pseudo-measurement signal over an additional cycle. Systematic underestimation of the burst peak persists in both EKF configurations since the single recursive pass cannot iterate the gradient signal to convergence.

\subsubsection{PLE with Time}
PLE rises sharply within the first ${\sim}10\,\mu$s of muon impact and plateaus by ${\approx}50\,\mu$s in both codes, because QP-induced error rates decay on the timescale governed by the trapping rate $s$ and diffusion coefficient $\kappa$; cycles beyond $50\,\mu$s operate near baseline and contribute minimally to the logical error budget. Therefore, PLE at $T = 100$ \textmu s serves as reasonable performance metric throughout. The relative ordering of all methods is stable across the full horizon, and the gap between genie and uniform-prior decoding opens entirely within the burst phase, confirming that the decoding advantage from accurate QP estimation is concentrated in the first few post-impact cycles.

\begin{figure}
    \centering    \includegraphics[width=0.95\linewidth]{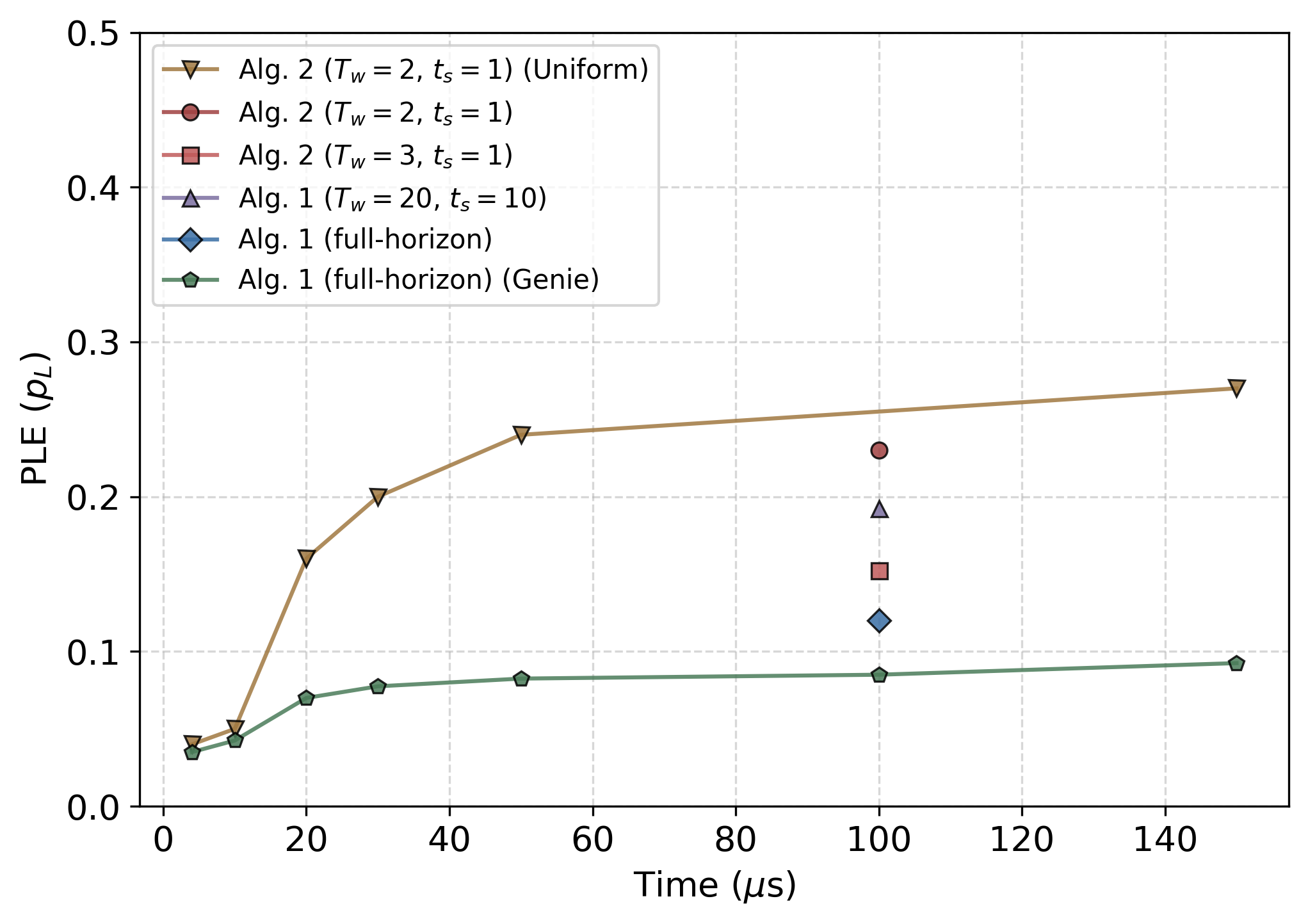}\vspace{-3mm}
    \caption{PLE vs. time for distance 7 surface code.}
    \label{fig:ple_vs_tslots_surface}
\end{figure}
\begin{figure}
    \centering
    \includegraphics[width=0.95\linewidth]{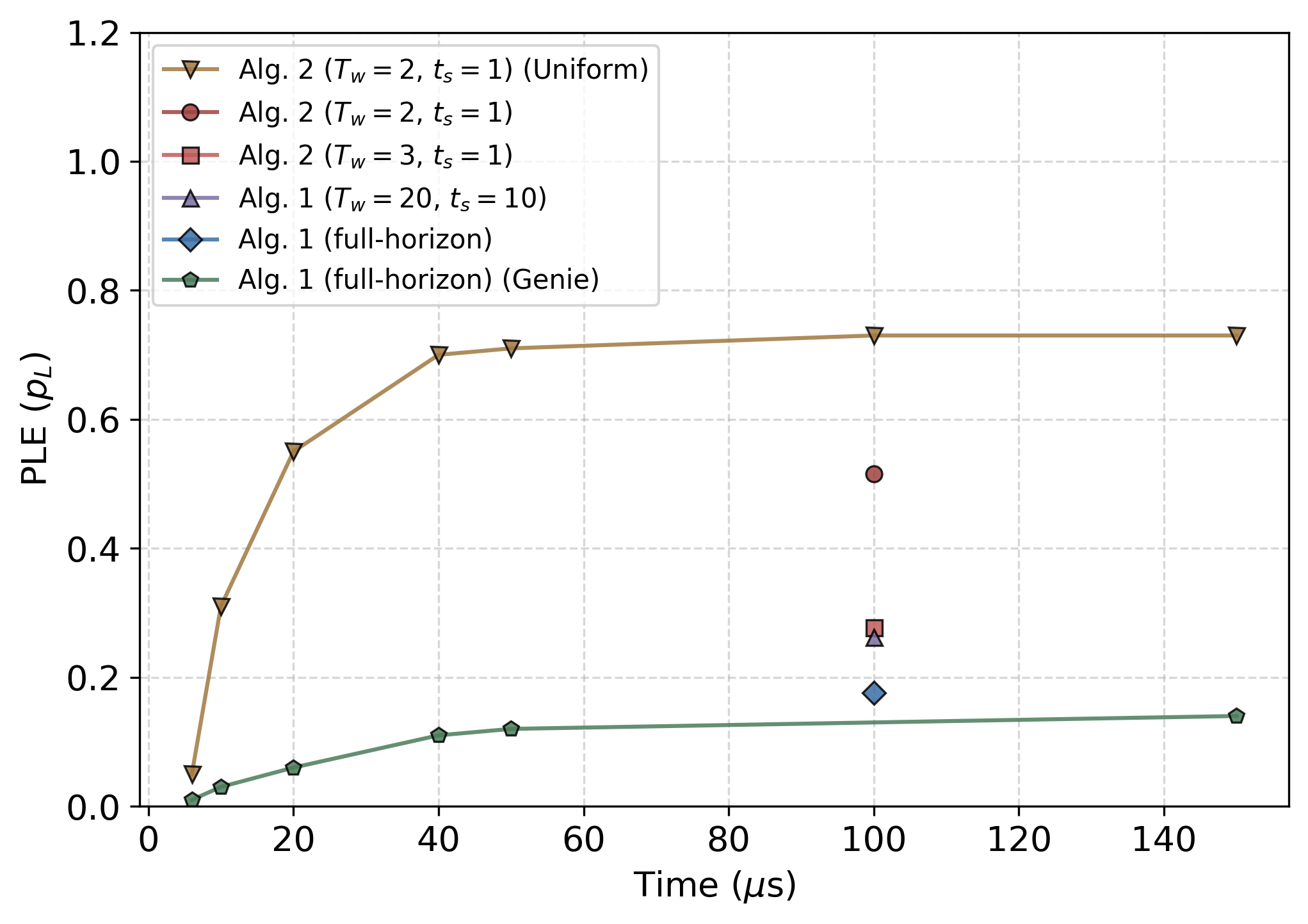}\vspace{-3mm} 
    \caption{PLE vs. time for [[72, 12, 6]] BB-qLDPC code.}
    \label{fig:ple_vs_tslots_bbqldpc}
\end{figure}

\subsubsection{Decoding Performance for Other Algorithm Configurations}
Alg.~\ref{algo:kalman} ($T_w = 3$, $t_s = 1$) achieves lower PLE than Alg.~\ref{algo:decode_sense_sliding_window} ($T_w = 20$, $t_s = 10$) on the surface code ($0.15$ vs.\ $0.19$) and comparable PLE on the BB-qLDPC code ($0.28$ vs.\ $0.26$), despite substantially worse QP density estimation. The logical error budget is dominated by a few of cycles immediately following muon impact, and the EKF's cycle-by-cycle update inflates error priors at the burst onset with minimal latency. The EKF avoids the latency penalty at the cost of spatial accuracy and elevated false alarms during the decay phase, but the decoding loss is likely more sensitive to missed detections of the burst than to false alarms during decay. This asymmetry induced by radiation-induced noise motivates Alg.~\ref{algo:kalman} as a preferred low-latency configuration.

\begin{figure}
    \centering
    \includegraphics[width=0.95\linewidth]{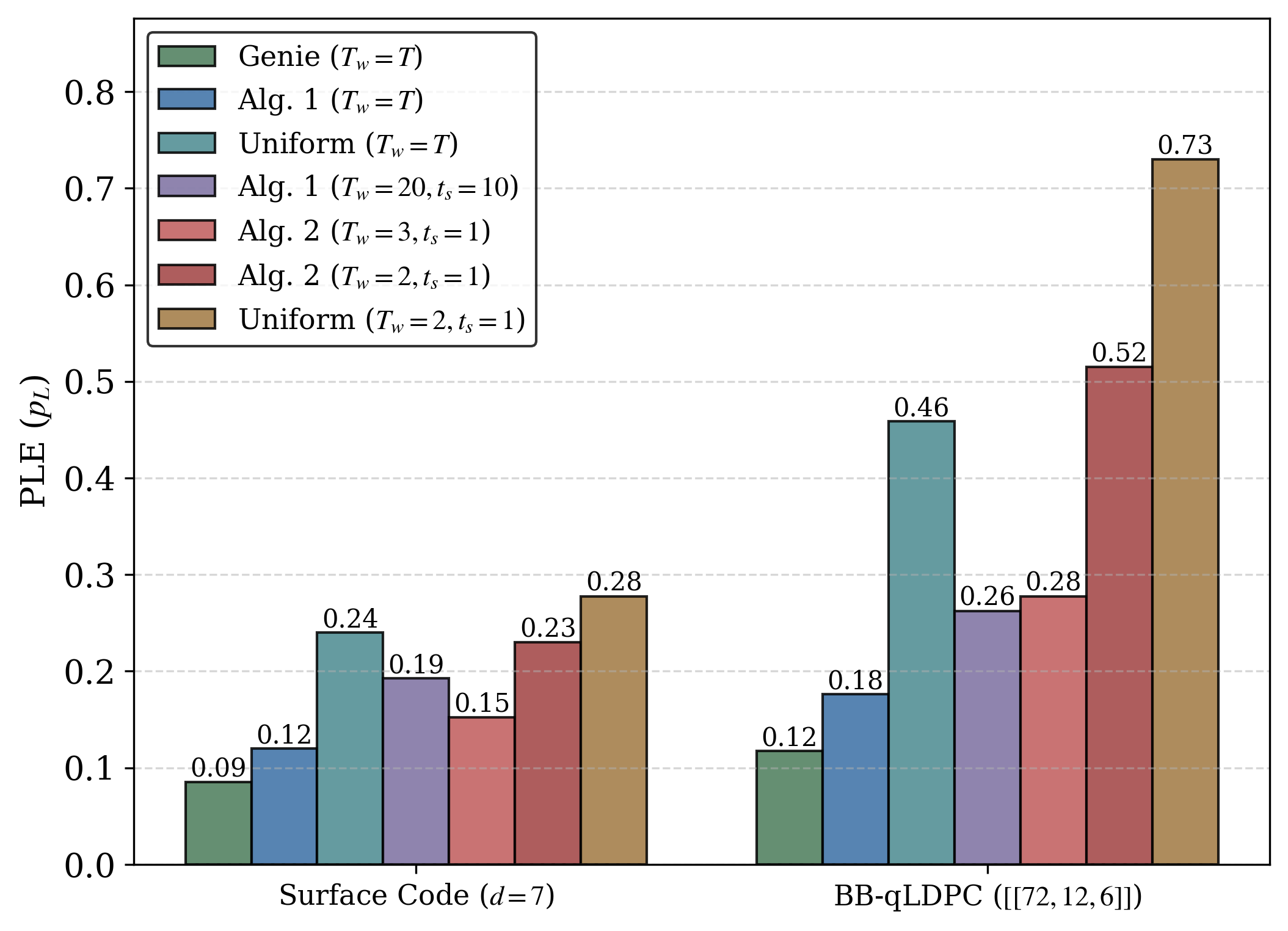}\vspace{-3mm}
    \caption{PLE at $100\,\mu s$ of distance 7 surface code and $[[72,12,6]]$ BB-qLDPC codes, using Alg.~\ref{algo:decode_sense_sliding_window}.}
    \label{fig:ple_vs_config_bar_graph_extra}
\end{figure}

\subsubsection{Convergence}
\label{apndx:convergence}
\begin{figure}
    \centering
    \includegraphics[width=0.95\linewidth]{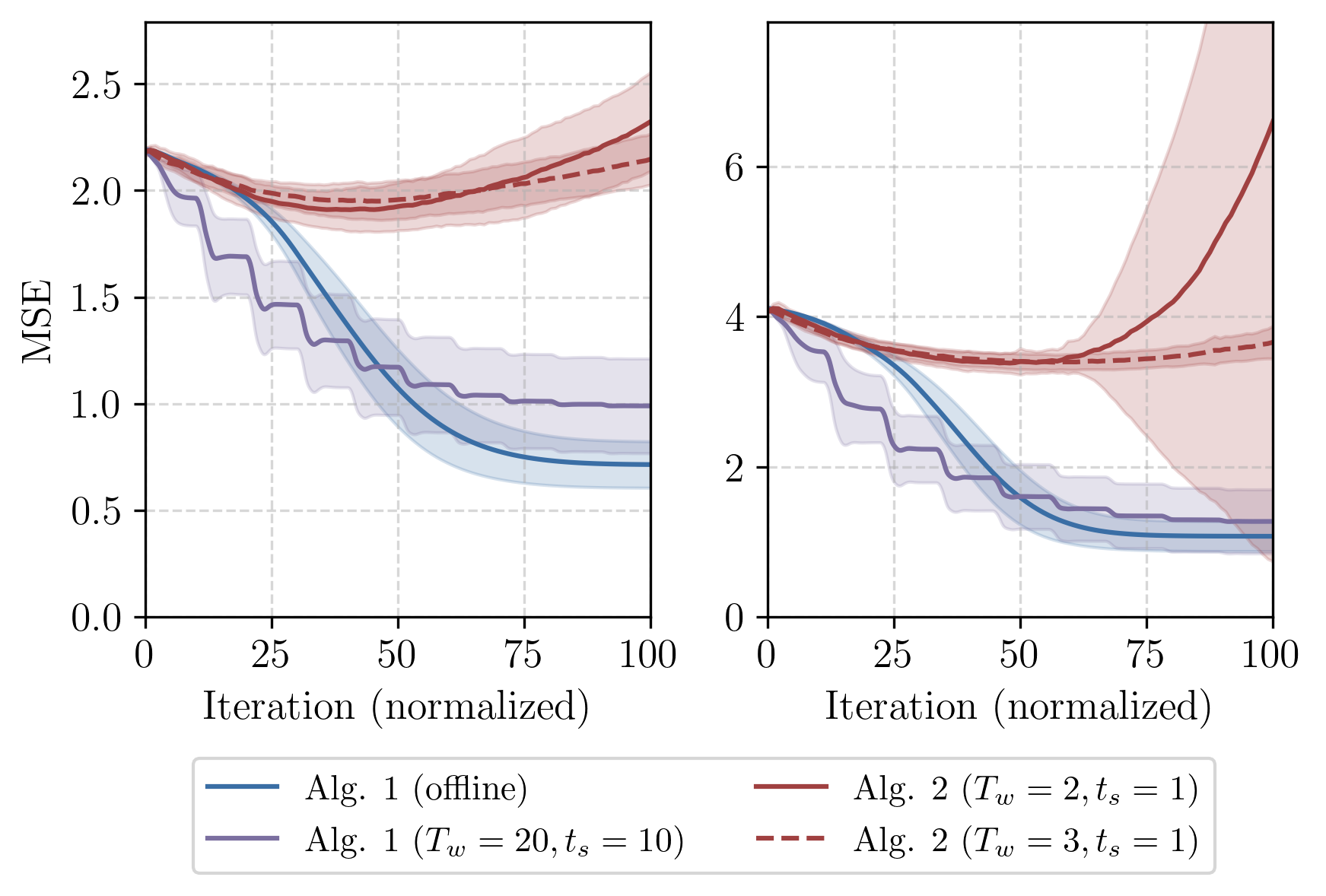}\vspace{-3mm}
    \caption{Evolution of MSE $=\frac{1}{JT} \sum_{t,i} (Z^{(k)}_{t,i} - \log X_{t,i})^2$ across iterations.}
    \label{fig:mse_history}
\end{figure}

Fig.~\ref{fig:mse_history} plots estimation MSE $= \frac{1}{JT}\sum_{t,i}(Z^{(k)}_{t,i} - \log X_{t,i})^2$ against iteration index.

Alg.~\ref{algo:decode_sense_sliding_window} (offline) converges smoothly and monotonically: the full-horizon window provides a globally consistent gradient signal, and backward coupling from post-burst syndromes coherently reinforces the burst-phase estimate at every iteration. The sliding-window variant exhibits periodic MSE jumps at window boundaries. MSE decreases within each window but converges to a higher floor than the offline variant.

Alg.~\ref{algo:kalman} ($T_w = 2$) shows an initial MSE decrease as the EKF detects the QP burst, followed by an increase during the decay phase. The latter arises from false alarms: the filter over-attributes residual syndrome activity to true hightened QP rather than statistical fluctuation. Extending to $T_w = 3$ reduces pseudo-measurement variance and lowers the decay-phase MSE, but the gap relative to Alg.~\ref{algo:decode_sense_sliding_window} persists across the full horizon.

\end{document}